\documentclass{article}      

\usepackage{epsfig}
\usepackage{natbib}
\usepackage{amssymb}
\usepackage{amsmath}
\usepackage{color}

\usepackage{graphicx}
\begin{document}

\title{Velocity derivatives in a high Reynolds number turbulent boundary layer.  Part I: Dissipation and Energy Balance}

%
\author{Michel Stanislas$^1$  \and Jean Marc Foucaut$^2$  \and  William K. George$^3$ \and Christophe Cuvier$^2$  \and Jean Philippe Laval$^2$}
\maketitle
\noindent$^1$ Pr. Emeritus, Centrale Lille, F59651 Villeneuve d'Ascq, France\\
$^2$ Univ. Lille, CNRS, ONERA, Arts et Metiers Institute of Technology, Centrale Lille,
UMR 9014 - LMFL - Laboratoire de M\'{e}canique des Fluides de Lille - Kamp\'{e} de F\'{e}riet,
F-59000 Lille, France\\
$^3$ Visiting Pr., Centrale Lille, F59651 Villeneuve d'Ascq, France
\medskip

\begin{abstract}
An experiment was performed using SPIV in the LMFL boundary layer facility to determine all the derivative moments needed to estimate the average dissipation rate of the turbulence kinetic energy, $\varepsilon = 2 \nu \langle s_{ij}s_{ij} \rangle$ where $s_{ij}$ is the fluctuating strain-rate and $\langle~\rangle$ denotes ensemble averages. Also measured
were all the moments of the full average deformation rate tensor, as well as all of the first, second and third fluctuating velocity moments except those involving pressure. The Reynolds number  was $Re_\theta = 7500$ or $Re_\tau = 2300$. 

The results are presented in three separate papers. This first paper (Part I) gives the measured average dissipation, $\varepsilon$ and the derivative moments comprising it.  It compares the results to the earlier measurements of \cite{balint91,honkan97} at lower Reynolds numbers and a new results from a plane channel flow DNS at comparable Reynolds number.  It then uses the results to extend and evaluate the theoretical predictions of ~\cite{george97b,wosnik00} for all quantities in the overlap region. Of special interest is the prediction  that  $\varepsilon^+ \propto {y^+}^{-1}$ for streamwise homogeneous flows and a nearly indistinguishable power law, $\varepsilon \propto {y^+}^{\gamma-1}$, for boundary layers. In spite of the modest Reynolds number, the predictions seem to be correct. It also predicts and confirms that the  transport moment contribution to the energy balance in the overlap region, $\partial \langle - pv /\rho - q^2 v/2 \rangle/ \partial y$ behaves similarly.  An immediate consequence is that the usual eddy viscosity model for these terms cannot be correct.

The second paper, part II \cite{george20},  examines in detail the statistical character of the velocity derivatives and identifies a particular problem with the breakdown of local homogeniety inside $y^+ = 30$.  A more general alternative for partially homogeneous turbulence flows is offered which is consistent with the observations. The details of the SPIV methodology necessary for processing velocity derivatives has been included as a separate paper, Part III \cite{foucaut20}, since it will primarily be of interest to experimentalists.

\end{abstract}

\section{Introduction}
\label{intro}

One of the most important parameters of modern turbulence theories and attempts to model turbulence is the average rate at which energy per unit mass, say $\varepsilon$, is dissipated through turbulence. It not only describes the actual dissipation of kinetic energy by deformation of the fluid elements~\cite{batchelor67}, it is also important to discussions of interscale energy transfer because of its relation to the spectral flux of energy from large to small scales \cite{george2014,TL72}.
Unfortunately the actual dissipation rate is also one of the most elusive quantities to obtain experimentally, mainly because the small scale motions (typically fractions of a {\it mm} in laboratory flows) and small spectral amplitudes which must be resolved to measure it correctly.

This paper extends the state-of-the art by combining measurements  in a moderately high Reynolds number turbulent boundary layer ($R_\theta = 7500$) using Dual Plane Stereo Particle Image Velocimetry (DP-SPIV)
and highly resolved Direct Numerical Simulation in a channel at comparable Reynolds number. The DNS is described in detail in section ~\ref{DNS-description}.
The experiment takes advantage of the unique boundary layer facility (Figure~\ref{fig-tunnel}) at Laboratoire de M\'{e}canique des Fluides de Lille (LMFL)  which can achieve boundary layers of up to 30 cm in thickness at Reynolds numbers up to $R_\theta = 20,000$ at speeds up to 10 m/s.  As will be demonstrated, the large length scales and low mean velocities allow state-of-the-art SPIV to resolve scales small enough to obtain reasonable estimations of the dissipation, while still insuring that the scattering particles follow the flow. But what really makes this experiment possible are two ``tricks'' used to reduce the SPIV quantization noise which would otherwise have prevented accurate derivative measurement. 

PIV typically has resolutions of less than 1:100 relative to the mean velocity, and this ``PIV noise'' is of order one (even much larger) when velocity differences at the scale of the Kolmogorov microscale are computed. When this noise is squared, it overwhelms the derivative. But by using crossed-planes (as shown in Figure~\ref{fig:cross-planevelocity}) all moments could be computed by multiplying the instantaneous quantities from SPIV in different planes where they intersect, but measuring the same flow.   The noise (from all sources) was largely uncorrelated (if not statistically independent), so the net noise contribution to the `measured' derivative moments was effectively zero, since no instantaneous quantity was squared.  The other ``trick'' was to use the continuity equation multiplied by one derivative and averaged so that all squared moments could be computed independently from only cross-moments. This could be used at all measurement locations, and provided an independent confirmations of the measurements along the intersecting planes.
The experiment and these techniques are described briefly below, and in detail in part III \cite{foucaut20}of this contribution.


\section{What is the dissipation?}

Before entering into the details of this contribution, it is important to note that many papers (even texts) in turbulence wrongly define this fundamental quantity. Therefore we begin with a careful review of the basic equations.

\subsection{The definition of dissipation}
The instantaneous rate of dissipation of kinetic energy, $\tilde{\varepsilon}$  (or simply {\it dissipation}) in any flow of any Newtonian fluid is defined to be:

\begin{equation}
	\tilde{\varepsilon} = 2 \nu \left[\tilde{s}_{ij}  \tilde{s}_{ij} - \frac{1}{3} \tilde{s}_{kk}\tilde{s}_{kk}\right]
\end{equation}
The tilde is used to represent an instantaneous quantity, $\nu$ is the kinematic viscosity, $\tilde{s}_{ij}$ is the instantaneous strain-rate defined by: 

\begin{equation}
	\tilde{s}_{ij} = \frac{1}{2} \left[ \frac{\partial \tilde{u}_i} {\partial x_j} + \frac{\partial \tilde{u}_j}{\partial x_i}\right]
\end{equation}
where $\tilde{u}_i$ is the instantaneous velocity. For an incompressible fluid flow like that considered herein,  $\tilde{s}_{kk}  = 0$, so the last term will be dropped in the remainder of this paper. The reason this is the true dissipation is NOT that the terms are all positive -- in fact 3 of them are usually negative. But it is this full $\tilde{\varepsilon}$ that shows up with negative sign in the entropy equation.  So it always acts to reduce the kinetic energy locally, and send it irreversibly to internal energy.

\subsection{The average dissipation rate.}

In turbulence studies, since the instantaneous flow is usually out of reach in practical conditions, it is the 
Reynolds averaged equations or spectral equations  which are solved.  So it is the average dissipation rate,

\begin{equation}
	\varepsilon \equiv \langle \tilde{\varepsilon} \rangle = \langle  2 \nu \tilde{s}_{ij}  \tilde{s}_{ij}\rangle
\end{equation}
that is of primary interest.  {\bf Hereafter in this paper the symbol, $\varepsilon$, will refer to the averaged dissipation rate, or simply ``the dissipation''.}

A common problem in turbulence, especially with DNS but also recently in reported experiments as well, has been the failure to recognize the correct definition of dissipation, and so report instead averaged values of  the `{\it pseudo-dissipation}',\footnote{Note that some refer to this as the `homogeneous dissipation', since $\mathcal{D} = \varepsilon$ when the flow is statistically homogenous. We consider this misleading, since it seems to imply that it is a dissipation in non-homogeneous turbulence, which it is not.  Only \cite{jakirlic02} appear to recognize this as a problem and correct for it.} 

\begin{equation}
	\mathcal{D} \equiv \nu \langle \left[ \frac{\partial u_i}{\partial x_j}\right]^2 \rangle,
\end{equation}
First note that $\tilde{\mathcal{D}} = \nu [\partial \tilde{u}_i / \partial x_j ]^2$ is NEVER the dissipation instantaneously.  The reason is of course that it contain both strain-rate and rotation-rate contributions, and the latter do not dissipate energy~\cite{batchelor67}.
But when averaged quantities are considered, the difference between $\varepsilon$ and $\mathcal{D}$ is often slight. This is due to the fact that high Reynolds number turbulence is often nearly {\it locally homogeneous}~\cite{george91}.  The reason is that in homogeneous flows $ \langle s_{ij} s_{ij} \rangle = \langle\frac{1}{2} \omega_i \omega_i \rangle = \langle \omega_{ij} \omega_{ij} \rangle$, where $\omega_i$ is the vorticity and $\omega_{ij}$ is the rotation-rate tensor~\cite{TL72, george91}.

But as will be shown in part II \cite{george20}, the tensorial versions $\varepsilon_{ij}$ and $\mathcal{D}_{ij}$ needed for Reynolds stress models differ near a solid surface (inside $y^+<30$). And probably near stagnation points and in separated flows as well.  In short, any time one velocity component is significantly reduced by the kinematic boundary condition.  Wrong values can lead to blaming the pressure-strain rate and turbulence diffusion models instead of the real culprit, a wrong dissipation  distribution among the components of the Reynolds stress balance equations. Appendix~\ref{sec-equations} details the source of the confusion and shows derivations of two forms of the turbulence kinetic energy equation: the one used by turbulence modelers involving $\mathcal{D}$, and the correct one  with the true dissipation, $\varepsilon$~\cite{TL72}.  

\subsection{Measuring the true dissipation}
To obtain the average dissipation rate,  $\varepsilon = 2 \nu \langle s_{ij} s_{ij} \rangle $  experimentally it is necessary to measure all its components. This in turn requires measuring the variance of each velocity derivative term together with a few covariances of them. There are a total of twelve terms which can be organized into three groups~\cite{george91}:
\begin{eqnarray}
	&\epsilon & =  \nu \left\{2\left[\left\langle \left[ \frac{\partial u_1}{\partial x_1} \right]^2 \right\rangle  + \left\langle \left[\frac{\partial u_2}{\partial x_2} \right]^2\right\rangle + \left\langle \left[\frac{\partial u_3}{\partial x_3} \right]^2 \right\rangle \right]\right. \nonumber \\ &  &  \hspace{.5in}
	+\left[ \left\langle \left[\frac{\partial u_1}{\partial x_2}\right]^2 \right\rangle  + \left\langle \left[\frac{\partial u_2}{\partial x_1}\right]^2 \right\rangle + \left\langle \left[\frac{\partial u_1}{\partial x_3} \right]^2\right\rangle \right.  \nonumber \\ & & \hspace{1in} +
	\left. \left\langle \left[ \frac{\partial u_3}{\partial x_1} \right]^2\right\rangle + \left\langle \left[\frac{\partial u_2}{\partial x_3} \right]^2 \right\rangle+\left\langle \left[\frac{\partial u_3}{\partial x_2} \right]^2\right\rangle \right]\nonumber \\ & &  \hspace{.5in} + ~
	2 \left[\left\langle \frac{\partial u_1}{\partial x_2}\frac{\partial u_2}{\partial x_1} \right\rangle + \left\langle \frac{\partial u_1}{\partial x_3}\frac{\partial u_3}{\partial x_1} + \left\langle \frac{\partial u_1}{\partial x_3}\frac{\partial u_3}{\partial x_1} \right\rangle\right] 
	\right\}
	\label{eq:02}
\end{eqnarray}
Note that the first nine are squared, but only the diagonals (top line) and crossed-moments (bottom line) have a factor of 2 in front of them.  The last line involves only cross-moments which are generally negative.

Obviously any attempt to determine the dissipation by experiment is especially difficult because of the large number of velocity derivatives which must be measured and the difficulty of measuring them in real turbulent flows.  As a result, all measurements to-date have had to make assumptions about the turbulence which are not necessarily valid. 


For the measurements preceding ours  reported here (all with hot-wires) it was difficult to measure accurately because of the high spatial resolution required \cite{wyngaard68,georgetaulbee92,ewing1995,ewing2000}. Only in the past few decades has it been possible to measure more than as few derivatives at a time, and even this required extraordinary efforts (c.f. \cite{balint91,honkan97,tsinober92,honkan97}).  All used Taylor's frozen field hypothesis for the streamwise derivatives. More recent attempts have been made using PIV, but most of these also used Taylor's hypothesis in some form.

For some recent PIV measurements (e.\ g.\ , \cite{carlier05,herpin11}), spatial resolution has been less of a problem than with hot-wires; but as noted above the limited dynamical range (typically less than 1:100) and the resulting noise present serious problems, since many  of the derivatives needed are squared.  
Time-resolved PIV has not  contributed  significantly for two reasons.  First of all, the data rates are simply not high enough yet to obtain a reasonable time derivative at the scales needed.  But second and more fundamentally,
Taylor's frozen field hypothesis has always been a serious issue when using time-derivatives (v. \cite{lumley65,georgeetal89,antonia80}).  The fundamental problem is that the convection velocity is fluctuating and this ``leaks'' energy down the derivative spectrum, so serious over-estimates of the derivative moments are common~\cite{champagne78}.

Even more serious problems with many measurements to-date arise from the need to make assumptions about the statistics (e.g., isotropy, homogeniety, etc.) to fill in the missing terms.  These are addressed in detail in paper II.


%

\subsection{The goal of this paper}

This paper will present our data and compare it to previous attempts to measure the dissipation and velocity derivatives in turbulent boundary layers. Of particular interest will be the data of
\cite{balint91} and \cite{honkan97} who published  experimental results of the dissipation rate obtained in a turbulent boundary layer with   specially designed multiwires probes \footnote{Note that the experiments of \cite{tsinober92} were of great interest as well, but we simply could not make their data fit our plots with the scaling parameters provided.}.  All of the DNS results for turbulent boundary layers (at least until very recently) are of much lower Reynolds number than the present experiment and therefore are of limited interest.  Only \cite{antonia91} examined the different hypotheses using DNS results from a channel flow at low Reynolds number.  Unfortunately many of the more recent DNS results which are at Reynolds numbers comparable to the present data did not catalog all of the necessary derivative moments to compute the true dissipation. For this reason, a specific channel flow DNS performed by \cite{thais11} was used for comparison here as all the necessary data where provided by the authors. The characteristics of this DNS are summarized in section~\ref{DNS-description}. It was expected to not be too different from a boundary layer in the near wall and log regions, nor was it.  

\begin{figure}[ht]
	\centering
	\resizebox{0.99\linewidth}{!}{\includegraphics[scale=1]{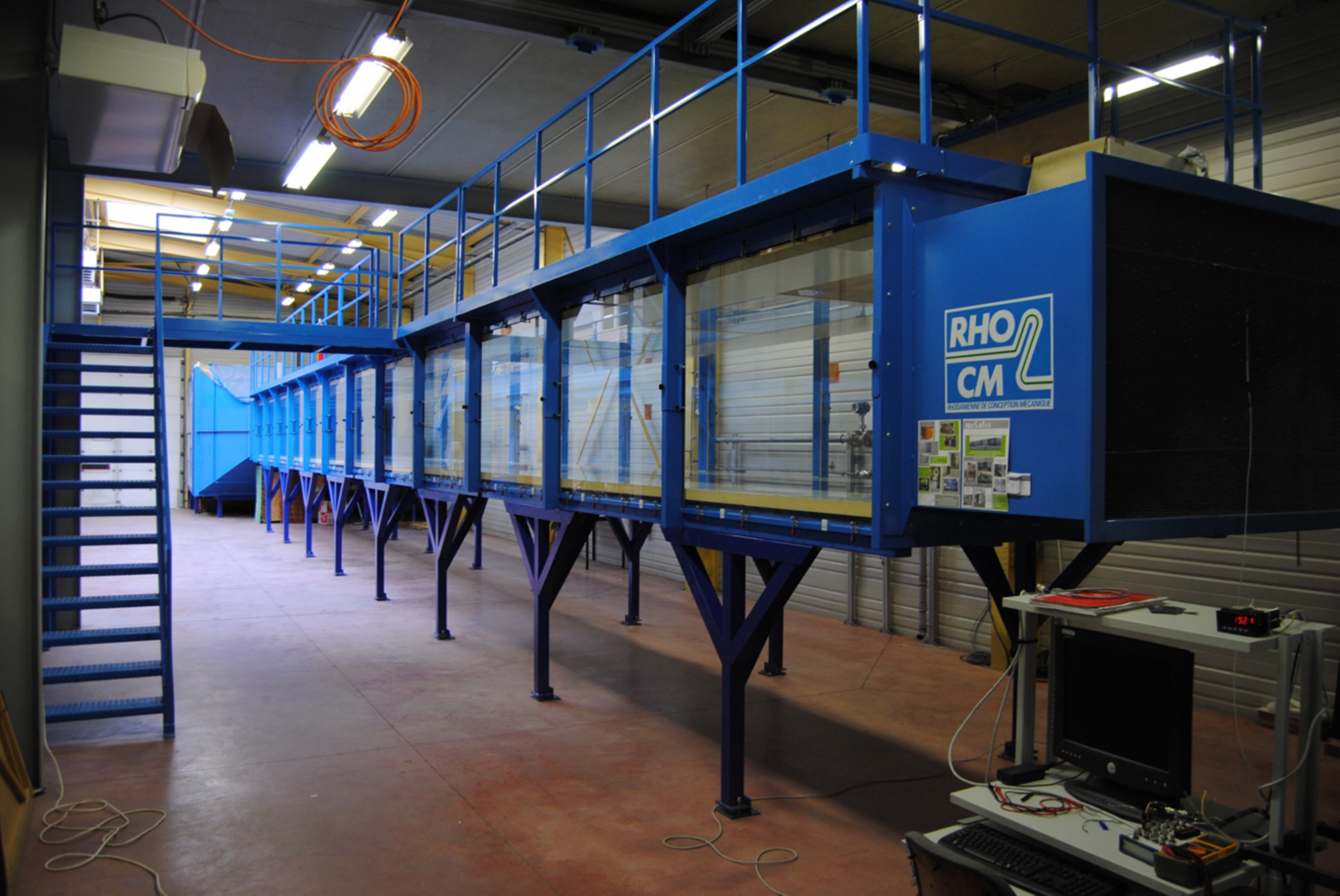}}
	\caption{Lille Boundary Layer Wind Tunnel showing 20 m test section with full optical access.}
	\label{fig-tunnel}      
\end{figure}

This Part I of our contribution presents the first measurements of {\it all} derivatives in a turbulent turbulent boundary layer at moderate Reynolds number using Dual-Plane-Stereoscopic PIV (or DP-SPIV) in the Lille Boundary Layer facility shown in Figure~\ref{fig-tunnel}.  SPIV is applied in two normal planes  to compute all the derivatives of the three velocity components in a turbulent boundary layer.
Both the true dissipation rate, $\epsilon$,  and the pseudo-dissipation rate, $\mathcal{D}$, are evaluated.
It also uses these new results and those of the channel flow DNS at comparable Reynolds number to evaluate previous measurements and theoretical predictions.

In part II \cite{george20} the assumptions usually made about the velocity derivatives and the dissipation (and enstrophy) inferred from them are examined in detail.  Also considered in part II \cite{george20} are the tensors, $\varepsilon_{ij}$ and $\mathcal{D}_{ij}$ which appear in  the Reynolds stress balance equations.
Finally in part III \cite{foucaut20}the DP-SPIV methodology, which should have wide application beyond this immediate paper, is analyzed in detail. It includes a detailed
discussion of the spatial resolution, the quantization noise, and previous suggestions for SPIV noise management using direct measurements. Also of interest should be how the continuity equation and the underlying statistical properties of the flow were employed to improve the accuracy of the results.

\section{Description of the channel flow DNS computation  \label{DNS-description}}
{\color{black}}

At the time of this work, there were no easily accessible DNS data sets which included the velocity derivative cross-velocity moment data.  (For example, even the recent relatively high Reynolds number data ~\cite{hoyas08} is missing these moments.)  Fortunately our colleagues (L. Thais and co-workers) made their $Re_\tau = 3000$ data available to us for post-processing. So the DNS results used in the present work were obtained by L. Thais and co-workers using a massively parallel code described in \cite{thais11} for the direct numerical simulation of Newtonian or viscoelastic turbulent channel flow. The spatial discretization uses Fourier modes in the two periodic streamwise and spanwise directions and six order compact finite differences in the normal direction. The Reynolds number based on friction velocity and half channel height is $Re_\tau=3000$. The discretized equations were integrated on a domain of size $8 \pi \times 2 \times 3 \pi/2$ with a spatial resolution $5120 \times 768 \times 2048$ in streamwise (x) normal (y) and spanwise (z) directions respectively. The first grid point in the normal direction is located at $\Delta_y^+=0.5$ from each wall and up to 18 points are used to discretize the first 10 wall units. The spatial discretizations in the streamwise and spanwise directions are $\Delta_x^+=11$ and $\Delta_z^+=7$ respectively. The simulation was performed on IBM Blue Gene/Q computer running at the IDRIS/CNRS computing center, Orsay, France. 

Up to 40 velocity fields were recorded but only the last 7 were used to compute the dissipation terms of the present analysis. These fields are separated by a non-dimensional time of 1.5 based on the bulk velocity and the half channel height. This corresponds to $\Delta t^+=200$ in wall unit time.  The dissipation terms were computed in physical space on the full domain using an $8^{th}$ order compact finite difference scheme for the derivatives in order to maintain accuracy.

\begin{figure}
	\centering 
	\includegraphics[width=0.99\columnwidth]{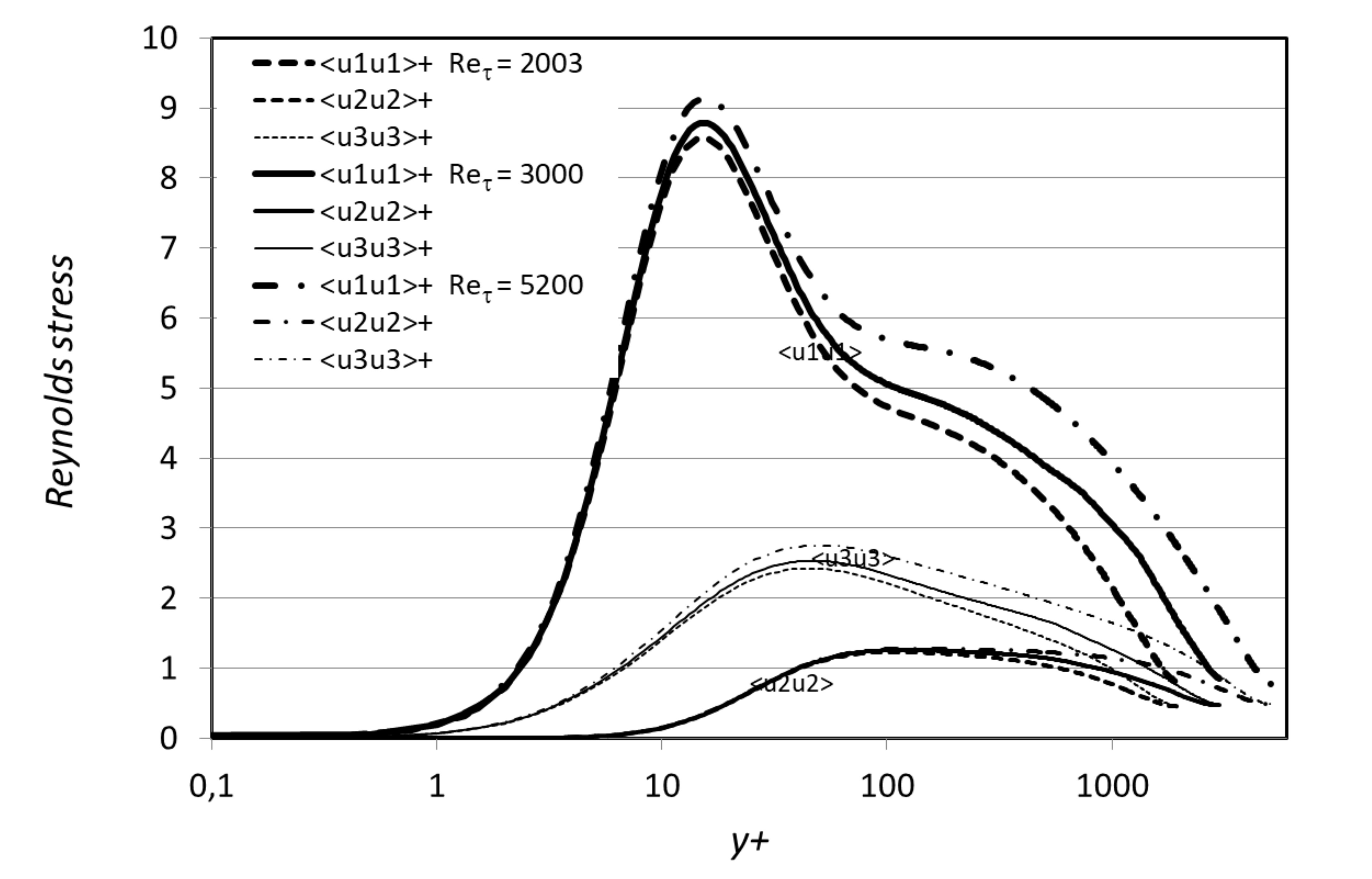}
	\caption{Comparison of the Reynolds stresses of the DNS of \cite{thais11} at $Re_\tau=3000$ with the DNS of \cite{hoyas06} at $Re_\tau=2003$, the DNS of  \cite{lee15} at $Re_\tau=5200$.}
	\label{fig:compare_dnsRS}
\end{figure}

Figure~\ref{fig:compare_dnsRS} shows a comparison of the Reynolds normal stresses of the present DNS with the DNS of ~\cite{hoyas06} at $Re_\tau=2000$ and the recent DNS of ~\cite{lee15} at  $Re_\tau=5200$. The three simulations agree very well in the near wall region. The first peak of $u^\prime$ which weakly depends on the Reynolds number lies within the two other DNS. The Reynolds number effect is clearly visible in the outer part, with the beginning of a plateau at the highest Reynolds number.

\begin{figure}
	\centering 
	\includegraphics[width=0.99\columnwidth]{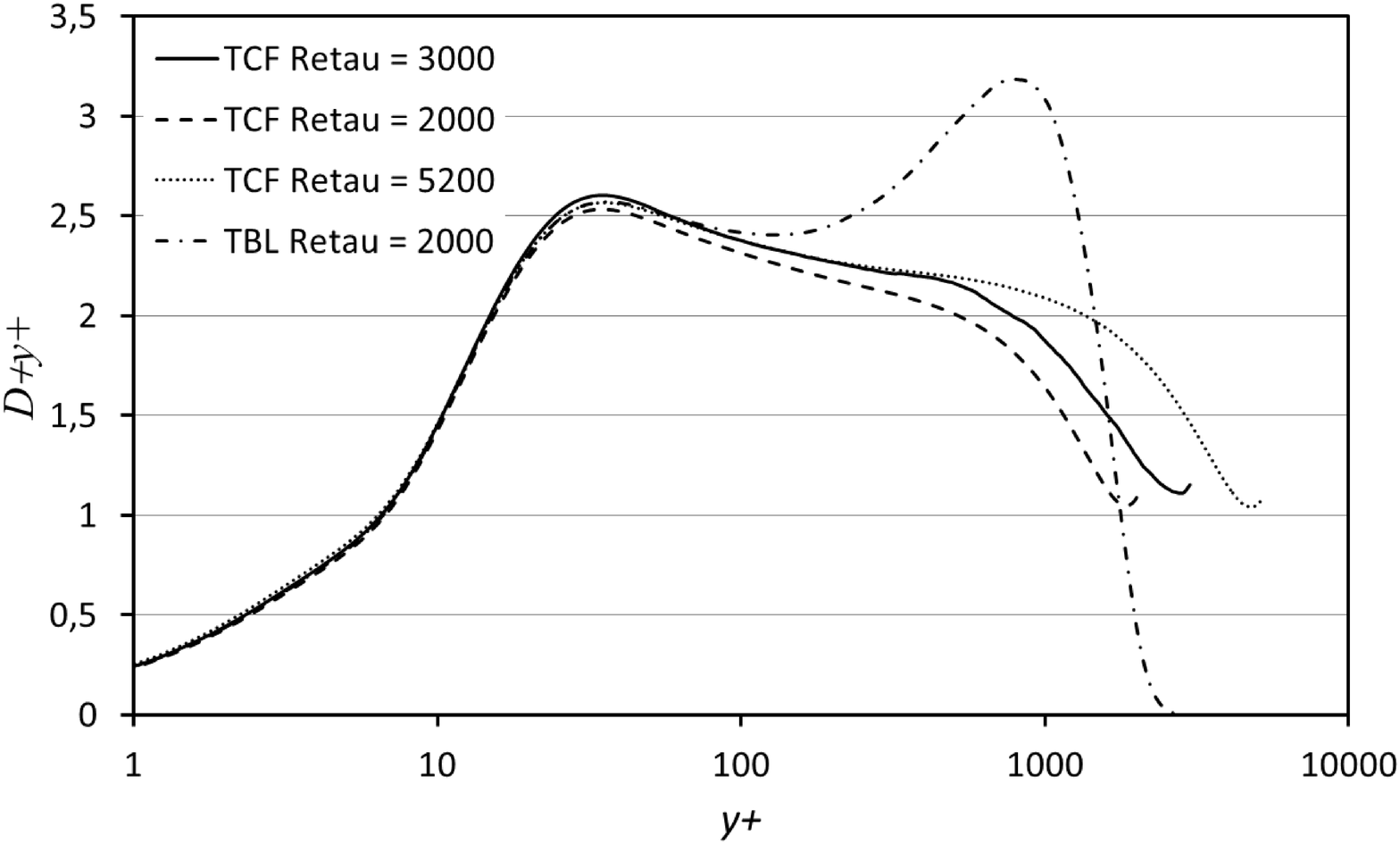}
	\caption{Comparison of the pseudo-dissipation  of the channel flow DNS of \cite{thais11} at $Re_\tau=3000$ with the DNS of \cite{hoyas06} at $Re_\tau=2003$, the DNS of \cite{lee15} at $Re_\tau=5200$. and the boundary layer DNS of \cite{sillero13}} 
	\label{fig:compare_dnsdissipation}
\end{figure}

Figure~\ref{fig:compare_dnsdissipation} shows the homogeneous dissipation $\mathcal{D}$ in inner variables along with that computed by Hoyas and Jimenez \cite{hoyas06} and ~\cite{lee15} for a plane channel and the data of \cite{sillero13} for a ZPG turbulent boundary layer. $\mathcal{D}$ was chosen here instead of $\varepsilon$ in order to compare the different data sets as not all of them provided the full dissipation. Also, the dissipation $\mathcal{D}^+$ has been premultiplied by the wall distance $y^+$. The reason for this representation will be justified in detail in the following but it can already be noted that it emphasises the Reynolds number influence in the outer part for the channel flow, the overlap region, and the difference between the channel and the boundary layer in the same outerpart.

\section{The experiment}
\label{sec-experiment}

The experiment was carried out in the Lille turbulent boundary layer wind tunnel shown in figure~\ref{fig-tunnel}. The optically accessible test section has been extended to the full tunnel length, and is 1 m high, 2 m wide and 20 m long.  The long length allows the development of the boundary layer so that its thickness  is close to 30 cm at 20 m.  The external flow has a slightly positive pressure gradient. The tunnel and flow quality has been documented in many previous papers~\cite{carlier05,stanislas08a,herpin11}.

\subsection{Stereo Particle Image Velocimetry}

Stereoscopic PIV (or SPIV) is now a recognized method for measuring turbulent flows. Many researchers have previously used this method to compute statistics of
the flow such as mean velocity, Reynolds stress tensor, probability density functions and even spectra (\cite{adrian00,foucaut11,herpin11}). SPIV allows the measurement of the three components of the velocity in a plane with an accuracy of about 1-2\% (0.1 pixel), but the limited spatial resolution (typically millimeters) cannot easily resolve the smallest scales of the flow.  To compute the dissipation components, both differentiation and statistical computations are involved and additional difficulties arise from the noise amplification when derivatives have to be computed from discrete realizations \cite{foucaut02}.  

In order to facilitate the presentation of the results we have chosen to include the detailed discussion of the measurement technique and especially the validation of it in a separate part III \cite{foucaut20}paper. Therefore we present only the briefest detail below.

\subsection{The dual plane SPIV setup}
The capabilities of the SPIV technique can be increased by the use of the light polarization to record velocity fields in two different planes simultaneously.
When the planes are parallel, this method is called dual plane stereoscopic PIV \cite{kahler00}.
The dual plane technique allows the measurement of two velocity fields with an adjustable time delay or spatial separation between them. By varying this delay, the space-time correlation of the velocity field can be computed. And by varying the separation, the 3D spatial correlation can be obtained. \cite{ganapathisubramani05a} used the dual plane technique to get the full gradient tensor and to study the near wall flow structures.

\begin{figure}[ht]
	\centering
	\resizebox{0.62\linewidth}{!}{\includegraphics[scale=1]{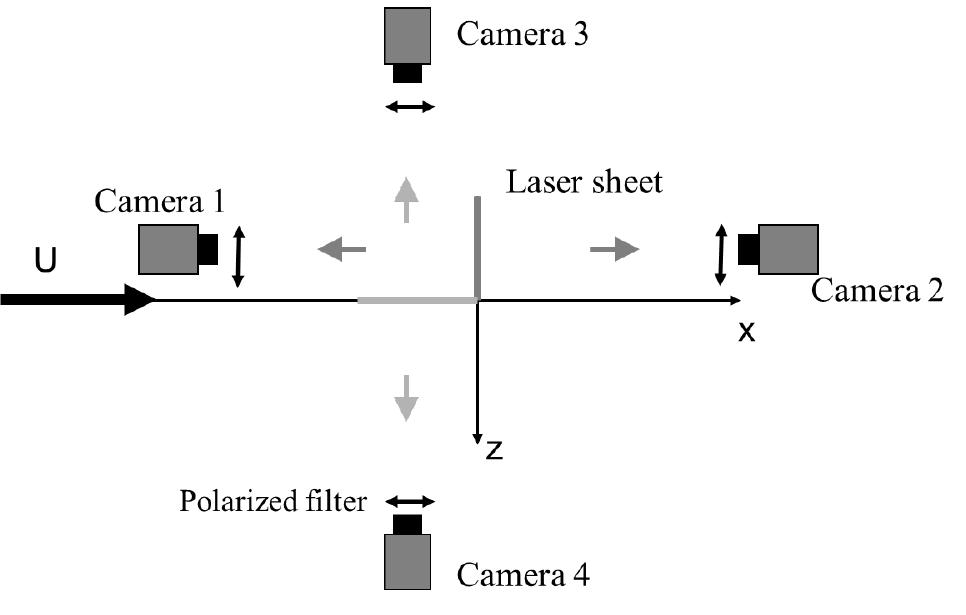}}
	\caption{Top view of the double SPIV experimental set-up}
	\label{fig-setup}      
\end{figure}

In the present experiment, the field of view was imaged with two Stereoscopic PIV systems in two normal planes (see Figure~\ref{fig-setup}).
The flow was seeded using Poly-Ethylene Glycol particles in the return circuit of the  tunnel just after the fan. 
The particle size was of the order of 1 $\mu m$ and the particle image size of the order of 1.8 pixel  on the cameras. Following \cite{raffel98}, such a size should lead to weak peak-locking, but none was apparent in the velocity pdf's shown in Part III \cite{foucaut20}.

The images from both cameras were processed with a standard multi-grid and multi-pass algorithm with image deformation \cite{scarano02}. The final interrogation window size was 24 x 32 pixels for each plane. Such a size corresponds to a square window in the physical space of 1.4 x 1.4 $mm^2$ (11.6 x 11.6 wall units). A mean overlap of about 66\% was used. The beam thickness varied from about 0.6 mm closest to the wall to 2 mm at the outermost location.  part III \cite{foucaut20}presents a detailed analysis of the spatial resolution effects on the measurements.

\subsection{Data processing and noise removal \label{sec-deriv-methodogy}}
The primary problem with using PIV for derivative measurement is the PIV noise.  For example, the derivative `signal' after processing, say $s(t)$, can be presented as $s(t) = d(t) + n(t)$ where $d(t)$ is the true derivative as a function of time, and $n(t)$ is the `noise' (or difference) between the true and measured signal. $n(t)$ can be thought of as a square-wave of random amplitude (see \cite{wanstrom07}).  So squaring and averaging yields $\langle s^2 \rangle = \langle d^2 \rangle + \langle n^2 \rangle$.
Note that we have assumed the noise to be uncorrelated with the signal.  (This is confirmed for our measurements in Part III \cite{foucaut20}).
For derivative measurement using SPIV, $\langle n^2 \rangle$ can be greater than $\langle d^2 \rangle$, even sometimes much greater.  

\begin{figure}[ht]
	\centering
	\resizebox{0.99\linewidth}{!}{\includegraphics[scale=1]{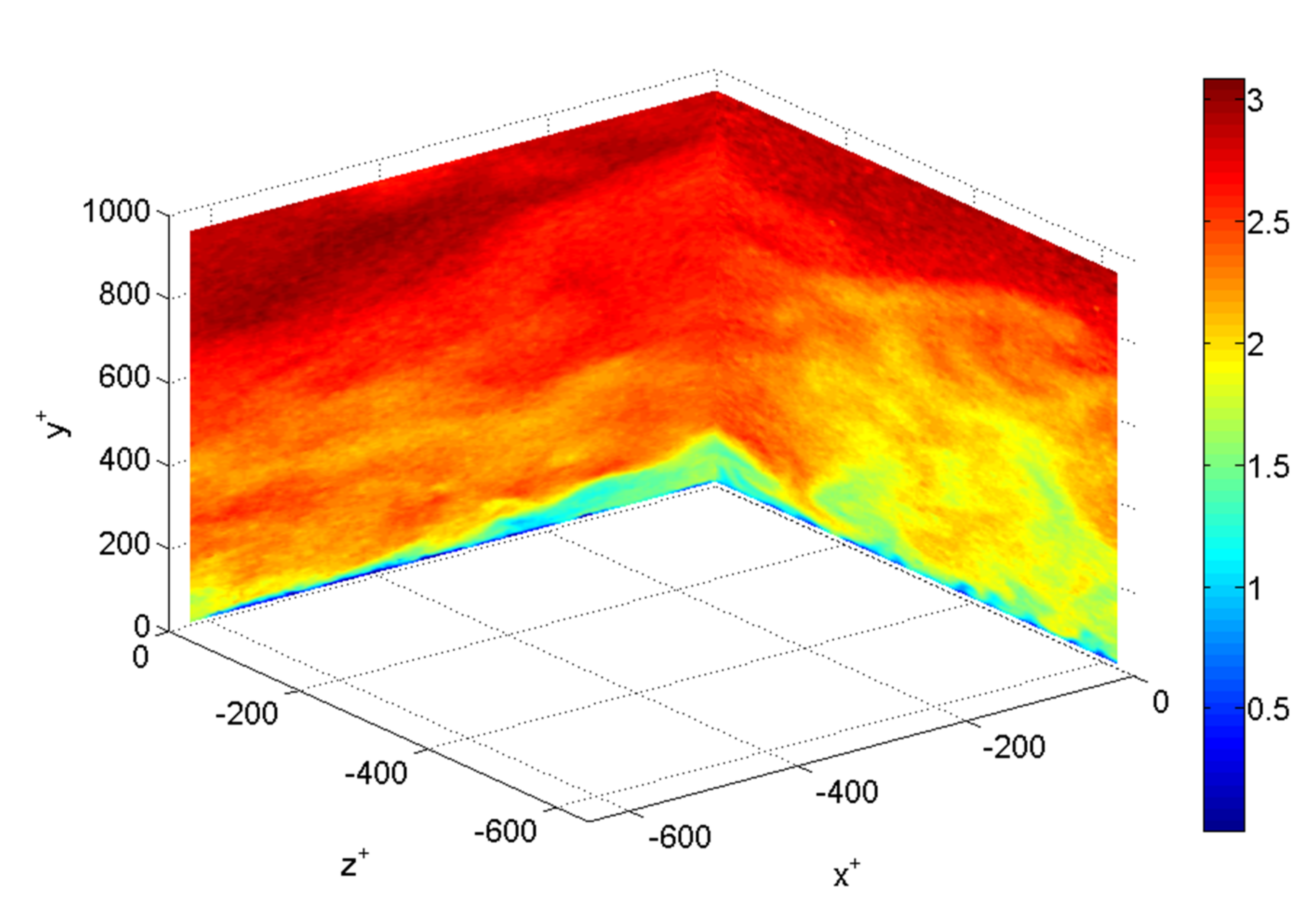}}
	\caption{Example of an instantaneous streamwise velocity field using cross-plane SPIV.  By using derivatives computed at the intersection of the two planes together with equation~(\ref{eq:cont}) it was possible to obtain most derivative moments without further assumptions (see Section~\ref{sec-experiment} and \cite{foucaut20} for details).}
	\label{fig:cross-planevelocity}      
\end{figure}

If, however, both planes are perpendicular and cross each other, as shown in Figure~\ref{fig:cross-planevelocity}, additional information about the spatial properties of the flow can be obtained~\cite{hambelton06,ganapathisubramani05b}.  But more importantly for the present experiment, it provides a unique opportunity to
evaluate noise-free derivative moment by choosing each term in the derivative product from different planes (since the noise is uncorrelated between the two planes).  For example, if $s_1(t)$ and $s_2(t)$ measure the same velocity but from different planes,  then their cross-correlation is given by:

\begin{equation}
	\langle s_1(t) s_2(t) \rangle = \langle d_1(t) d_2(t) \rangle + \langle n_1(t) n_2(t) \rangle
\end{equation}
But $\langle d_1(t) d_2(t) \rangle = \langle d(t)^2 \rangle$ if they measure the same (or nearly the same) interrogation volume.  And since the noise sources are statistically independent (or at least uncorrelated), $\langle n_1(t) n_2(t) \rangle = 0$.  This eliminates a primary source of noise in the estimation of the dissipation since almost all the derivative moments are squared quantities.  

The other ``trick'' used was to employ the continuity equation multiplied by one of the derivatives of interest, i.e.,

\begin{equation}
	\frac{\partial u_m}{\partial x_n} \left[ \frac{\partial u_1}{\partial x_1} + \frac{\partial u_2}{\partial x_2} + \frac{\partial u_3}{\partial x_3}\right]  = 0
	\label{eq:cont}
\end{equation}
Setting $m,n = 1$ for example yields:

\begin{equation}
	\left[\frac{\partial u_1}{\partial x_1}\right]^2 = - \left[\frac{\partial u_1}{\partial x_1}\frac{\partial u_2}{\partial x_2} + \frac{\partial u_1}{\partial x_1}\frac{\partial u_3}{\partial x_3}\right] 
	\label{eq:cont11}
\end{equation}
So all of the diagonal terms in the dissipation tensor can be computed from `noise-free' cross-moments. This proved particularly useful for confirming the validity of the other assumptions and techniques used in Part III \cite{foucaut20}. This is especially true when used together with the assumption of {\it local homogeneity} which as shown in part II \cite{george20} is valid over most of our flow (outside of $y^+ = 30$). 
Finally derivatives were also computed using the `de-noising' methodology of ~\cite{foucaut02}, which we tested and validated in Part III \cite{foucaut20}.

\subsection{Temporal and spatial resolution}

\begin{figure}
	\resizebox{0.85\linewidth}{!}{\includegraphics[scale=1]{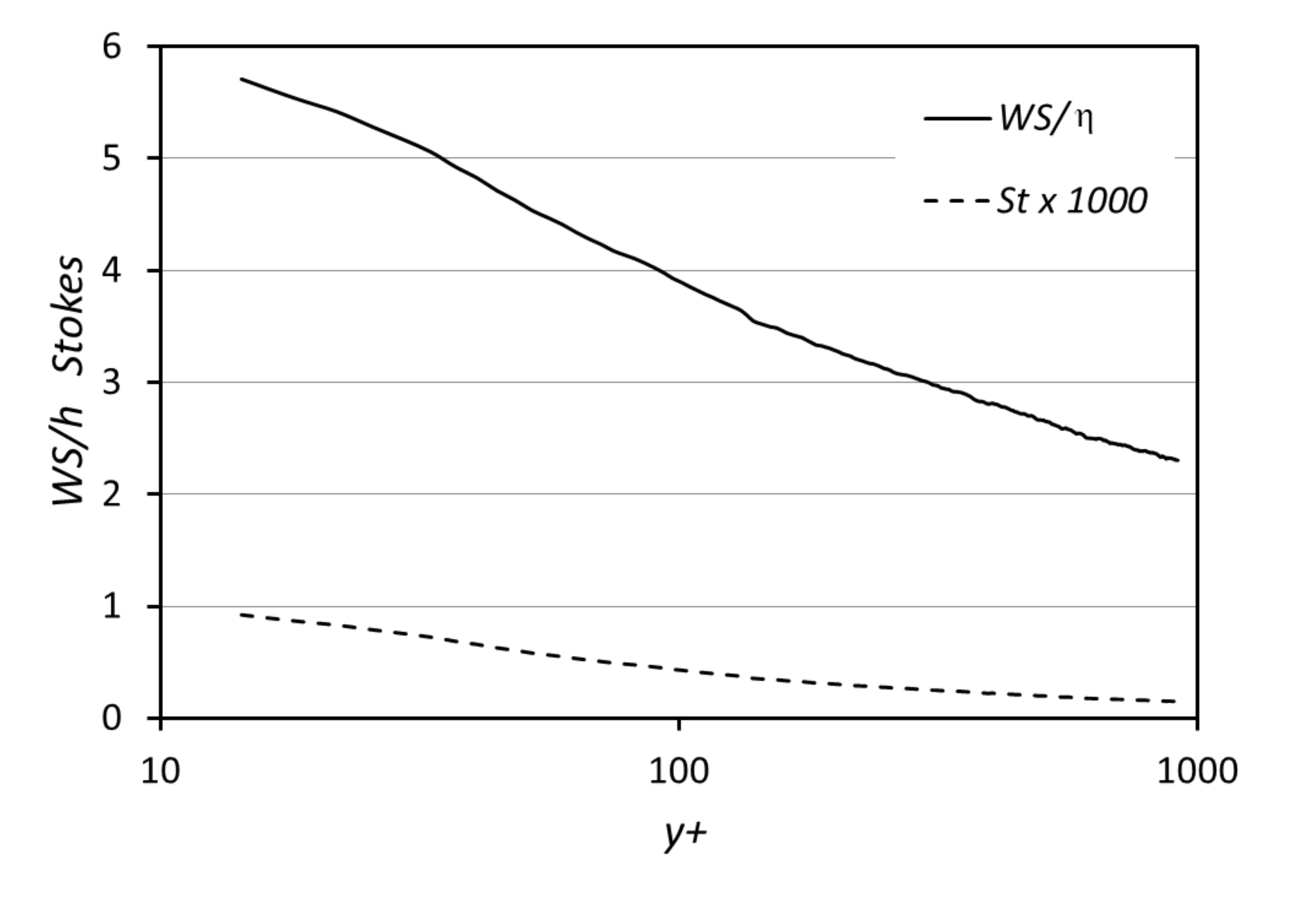}}
	\caption{Interrogation window size normalized by the Kolmogorov microscale, $(\nu/\varepsilon)^{1/4}$ and particle time scale normalized by the Kolmogorov microtime scale, $(\nu/\varepsilon)^{1/2}$. Wall distance $y$ normalized by $\nu/u_\tau$ using the data below.}
	\label{fig:Spacetimeresolution}      
\end{figure}

Figure \ref{fig:Spacetimeresolution} shows that the time-scale ratio for the scattering particles (sometimes referred to as Stokes number, $St^+$) is much less than one. So indeed the particles contributing to the velocity in the interrogation volume can be assumed to follow the flow at the dissipative scales. 

Also shown in the figure~\ref{fig:Spacetimeresolution} is the ratio of largest dimension of the SPIV interrogation window size, say $WS$, to the Kolmogorov microscale, $\eta = (\nu^3/\varepsilon)^{1/4}$, estimated from the data. The values range from $2$ at the outermost points of our measurement to $6$
at the innermost.  These are almost identical to those reported
by~\cite{balint91,honkan97}, suggesting any differences cannot be attributed to spatial resolution. Here, thanks to the overlap of 66\% of the interrogation windows (which gives a grid point in the data every $4 \eta$ at worst, with 33\% of new information in the interrogation window), one can expect an effective spatial resolution slightly better than the interrogation window size $WS$, that is varying between 5 and 1 $\eta$ from the wall outward.

Since about 99\% of the dissipation lies at spectral wavenumbers $k$ below $k \eta =1$, it requires  $WS \le \pi \eta$ \cite{georgetaulbee92} to resolve the entire derivative spectrum. But since $\varepsilon$ is an integral under the dissipation spectrum which has a long exponential tail, somewhat larger ratios are indicated to be acceptable~\cite{wyngaard68,ewing1995,ewing2000}. So for most of our flow spatial filtering should not be an issue, except perhaps at the closest wall positions.  The whole subject of spatial filtering is of considerable interest, and is addressed in detail in part III \cite{foucaut20}and in the conclusions of the present paper. 

\section{Experimental Results \label{sec:experimentalresults}}
In the present paper,  Dual-Plane-Stereoscopic PIV in two normal planes is applied to compute all the derivatives of the three velocity components in a turbulent boundary layer using the techniques briefly described above. (As noted before, these techniques are discussed in detail and validated in part III \cite{foucaut20}of the present contribution.)  

{\color{black} This section presents the measured derivative moments and compares them to earlier measurements and DNS.  They represent the principal results of this investigation.  These data are then used in Section~\ref{sec-asymptotictheory} to characterize the turbulence energy dissipation in the overlap and the near wall regions.  
	
	\subsection{Experimental parameters and velocity moments}
	
	The present experiment was carried out at a Reynolds numbers $R_\theta$ of 7500 ($\delta^+$ of 2300) which corresponds to a free stream velocity of 3  m/s. Table ~\ref{table-parameters} summarizes the important boundary and integral parameters for this experiment.
	
	\begin{table}
		\caption{Table showing experimental flow parameters}
		\label{table-parameters}       
		\begin{center}
			\begin{tabular}{llllllll}
				\hline\noalign{\smallskip}
				$Ue$ & $\delta$ &  $\delta*$ & $\theta$ & $u_\tau$ & $Re_\theta$ & $Re_\tau$ & $C_f$  \\
				\noalign{\smallskip}\hline\noalign{\smallskip}
				3 m/s & 0.32 m & 48.2 mm & 36.2 mm & 0.113 m/s & 7634 & 2598 & 0.00275\\
				\noalign{\smallskip}\hline
			\end{tabular}
		\end{center}
	\end{table}

	Figures ~\ref{fig:Mean_Velocity_Profile} show the mean streamwise velocity measured from the two PIV planes, compared with hot wire anemometry (HWA). Figure~ \ref{fig:Turb_Int_Profiles}  gives the  turbulent intensity profiles measured from the same two PIV planes,  compared with hot wire anemometry (HWA) and DNS data. These results are in very good agreement above 15 wall units, except for the spanwise turbulence intensity component $\sqrt{\langle w^2 \rangle}$ from the hot wire  in figure \ref{fig:Turb_Int_Profiles}. These data obviously depart from the PIV and the DNS below 40 wall units. This can be attributed to hot-wire errors resulting from the mean velocity gradient at the scale of the X-wire probe and cross-flow errors from the increasing local turbulence intensity near the wall, neither of which can be taken into account by calibration. The XY data are a bit lower than the YZ  and DNS ones as the $w$ component is out of plane in the XY plane. Also, a slight discrepancy is observed between the two PIV planes around the peak of $\sqrt{\langle u^2 \rangle}$. The XY plane data are quite low at one single point at $y^+ \simeq 15$. A careful check showed that one of the PIV cameras had a set of burned-out pixels in that region, which explains the discrepancy. Finally, also close to the wall, $\sqrt{\langle v^2 \rangle}$ is in good agreement between the PIV and hot wire data, but departing from the DNS when approaching the wall. This can be attributed to a lack of spatial resolution in both cases, agravated by the fact that this is the smallest component of the three. Beside these very near wall problems, the agreement is quite good between the two PIV planes the HW and the DNS.

\begin{figure}
	\centering 
	\resizebox{0.95\linewidth}{!}{\includegraphics[scale=1]{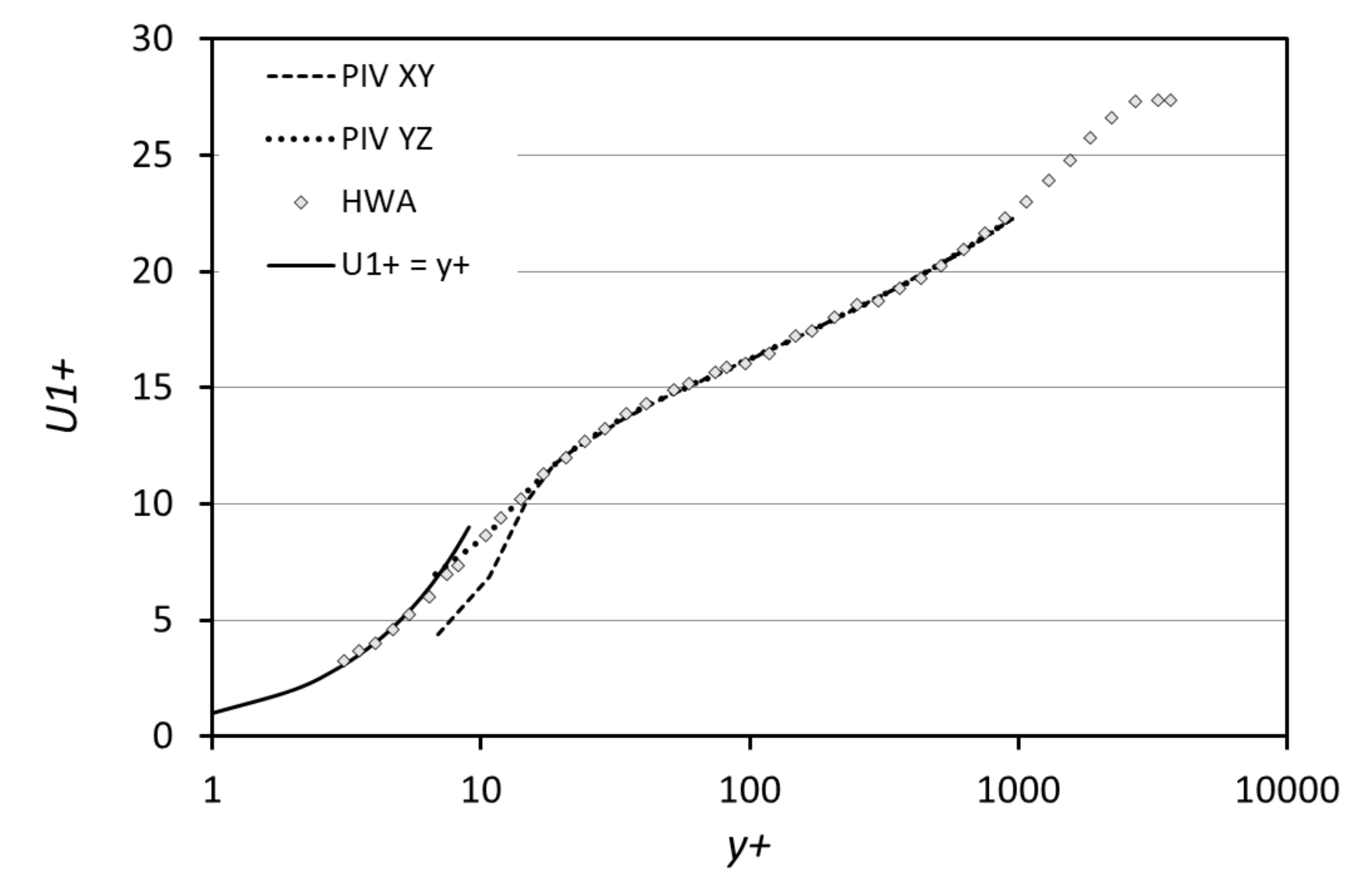}}
	\caption{Mean velocity profiles of the two PIV planes compared to hot-wire data.}
	\label{fig:Mean_Velocity_Profile}      
\end{figure}

\begin{figure}
	\resizebox{0.95\linewidth}{!}{\includegraphics[scale=1]{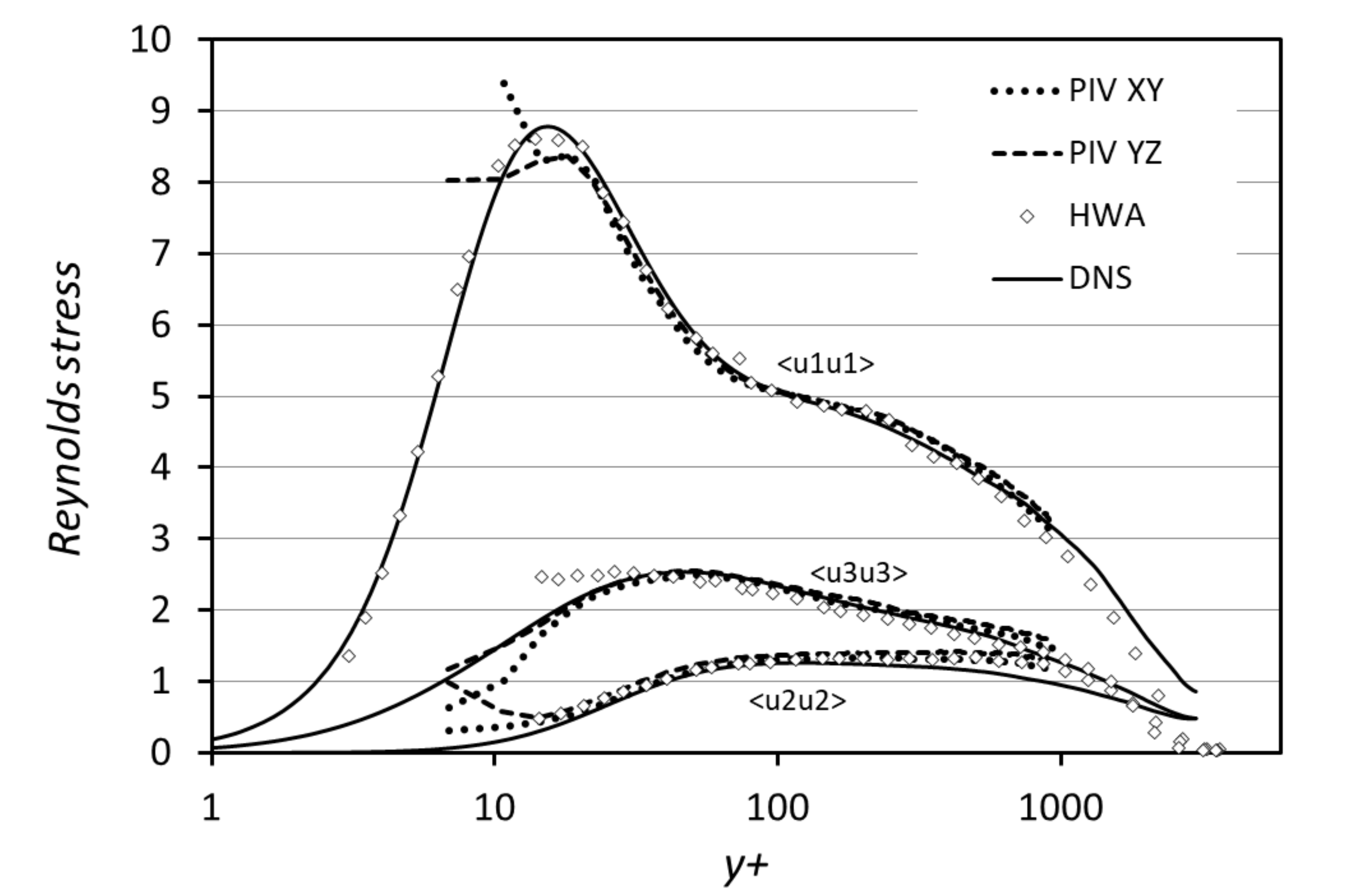}}
	\caption{Turbulence intensity profiles of the two PIV planes compared to hot-wire  and DNS data.}
	\label{fig:Turb_Int_Profiles}      
\end{figure}	
	
	\subsection{Measured derivative moments \label {sec-derivmoments}}
	
	The derivatives were computed by the methods described briefly in Section~\ref{sec-deriv-methodogy}, and in detail in part III \cite{foucaut20} of this paper. Based on the accuracy analysis performed in Part III \cite{foucaut20}, the data are supposed to be valid to within a few percent  for $y^+> 25$, indicated by the vertical dashed line on the plots. But as will be seen, this analysis does not take into account the spatial filtering. The term $\langle (\partial u_3 / \partial x_1) (\partial u_1/\partial x_3) \rangle$ was computed only at the intersection of the two SPIV planes and cannot therefore be averaged along one of the two planes, so it shows more scatter than the other profiles.  Table ~\ref{tab:measurementsummary} summarizes how each derivative moment was measured and which corrections were applied.
	
	\begin{table}
		\caption{Summary showing how each derivative moment was obtained}
		\label{tab:measurementsummary}       
		\begin{center}
			\begin{tabular}{lllll}
				\hline\noalign{\smallskip}
				$\left\langle \frac{\partial u_i}{\partial x_j}\frac{\partial u_i}{\partial x_j} \right\rangle$ & plane &  cross-plane & noise-corrected & continuity   \\
				\hline \noalign{\smallskip}
				i = 1 j = 1 & XY & no & yes & yes\\
				i $\neq$ 1 j = 1 & XY & no & yes & no\\
				i = 2 j = 2 & both & yes & yes & yes\\
				i $\neq$ 2 j = 2 & both & yes & yes & no\\
				i = 3 j = 3 & YZ & no & yes & yes\\
				i $\neq$ 3 j = 3 & YZ & no & yes & no\\
				\noalign{\smallskip}\hline
			\end{tabular}
		\end{center}
	\end{table}

	Figures~\ref{fig:allderivatives-lin-lin} to \ref{fig:allderivativestimesy-lin-log} show profiles of all twelve derivative moments contributing to the true dissipation $\varepsilon$ (Eq. [\ref{epsilon}]) in different formats: linear-linear, linear-log and linear-log premultiplied by $y^+$. 
	Each plot displays different features of the data which will be discussed below. But all plots make it clear that the derivative moments are not all the same.  In particular, it should be noted that all cross products are negative. The questions of whether the derivative moments are {\it locally homogenous}, {\it locally axisymmetric} or {\it locally isotropic} will be addressed in detail in part II \cite{george20}, together with the question of their individual contribution to the total dissipation. It should be obvious that the {\it locally isotropic} hypothesis is highly unlikely to be satisfactory. And the near wall region will be extremely problematic as well for all hypotheses. 
	Note that  even though many moments peak strongly near the wall, some go to zero at $y=0$. Note  also that none of our measurements go to zero at large values of $y$ since measurements were not taken outside of the overlap (or log) region.
	
	\begin{figure}
		\resizebox{1.0\linewidth}{!}{\includegraphics[scale=1]{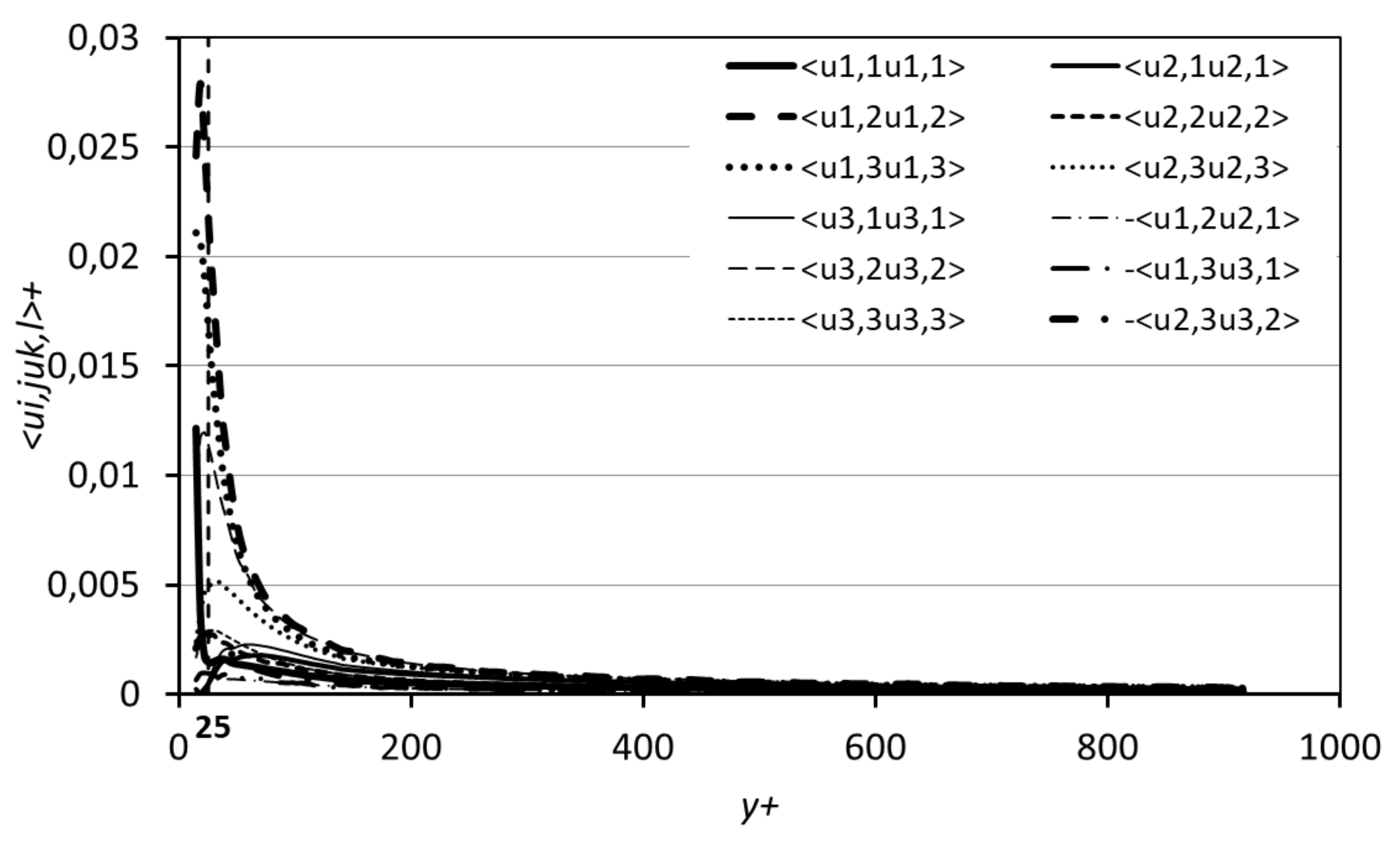}}
		\caption{Linear-linear plot of all dissipation derivatives moments deduced from the PIV data plotted together.}
		\label{fig:allderivatives-lin-lin}      
	\end{figure}

	Figure~\ref{fig:allderivatives-lin-lin} clearly evidences the difficulty of measuring accurately the dissipation in the outer part and even the overlap region due to the small values of the different components. It will be seen that noise is really an issue here. In the very near wall region, the values of the terms are higher but spatial resolution becomes a real issue for all experimental techniques.
	
		\begin{figure}
		\resizebox{1.0\linewidth}{!}{\includegraphics[scale=1]{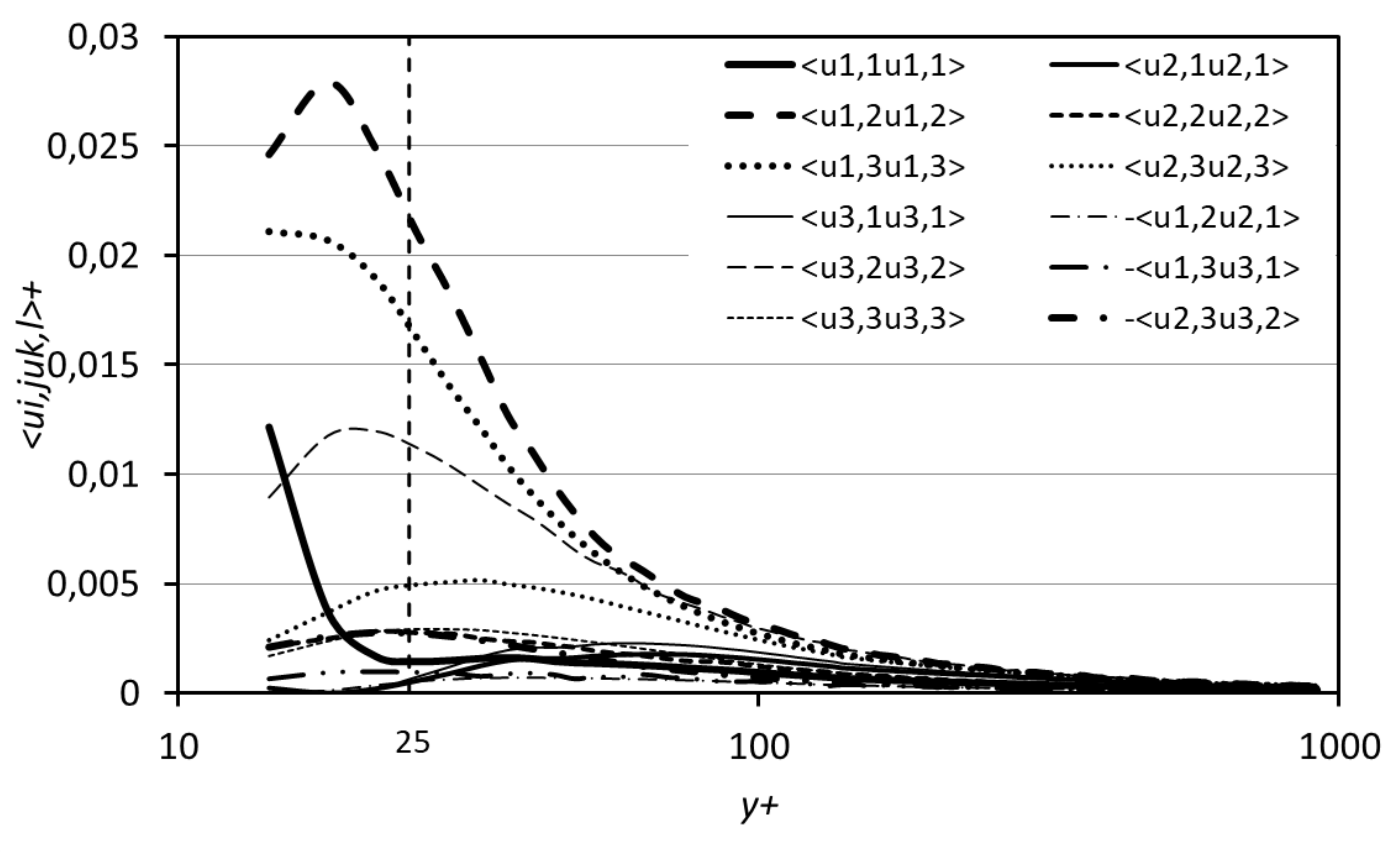}}
		\caption{Linear-logarithmic plot of all dissipation derivatives moments deduced from the PIV data plotted together.}
		\label{fig:allderivatives-lin-log}      
	\end{figure}

	Figure~\ref{fig:allderivatives-lin-log} emphasizes the large differences between the different components in the very near wall region (ratios of 10 or even more are observed near $y^+ = 20$). This has an interesting experimental consequence: it is probably not necessary to measure all the components of $\varepsilon$ to have a good estimation of it. The consequences on the dissipation tensor $\varepsilon_{ij}$ will be looked at in part II \cite{george20}.
	Figure~\ref{fig:allderivativestimesy-lin-log} sheds already some light on what will be a key result of the present contribution: the fact that the dissipation terms behave as $1/y^+$ in the overlap region.
	\cite{george97b,wosnik00} argue from Near-Asymptotics that the dissipation in the overlap region of a channel or pipe flow varies as $1/y$; i.e., $\varepsilon^+ = D / y^+$, where $D$ depends weakly on $\ln \delta^+$ and is asymptotically constant.  The result is exact for a parallel flow homogeneous in the streamwise direction.  \cite{george97b} argue it is at least approximately true for zero-pressure gradient boundary layers even if it varies as a power law $\varepsilon^+ \propto D_i/{y^+}^{1-\gamma}$ since $\gamma$ is a small parameter which decreases with increasing $\ln \delta^+$~\cite{george97b}. Note that the energy balance in the overlap region implies this must be the same $\gamma$ as for the mean velocity.  Some background for this is provided in Appendix B. 
	
	\begin{figure}
		\resizebox{1.0\linewidth}{!}{\includegraphics[scale=1]{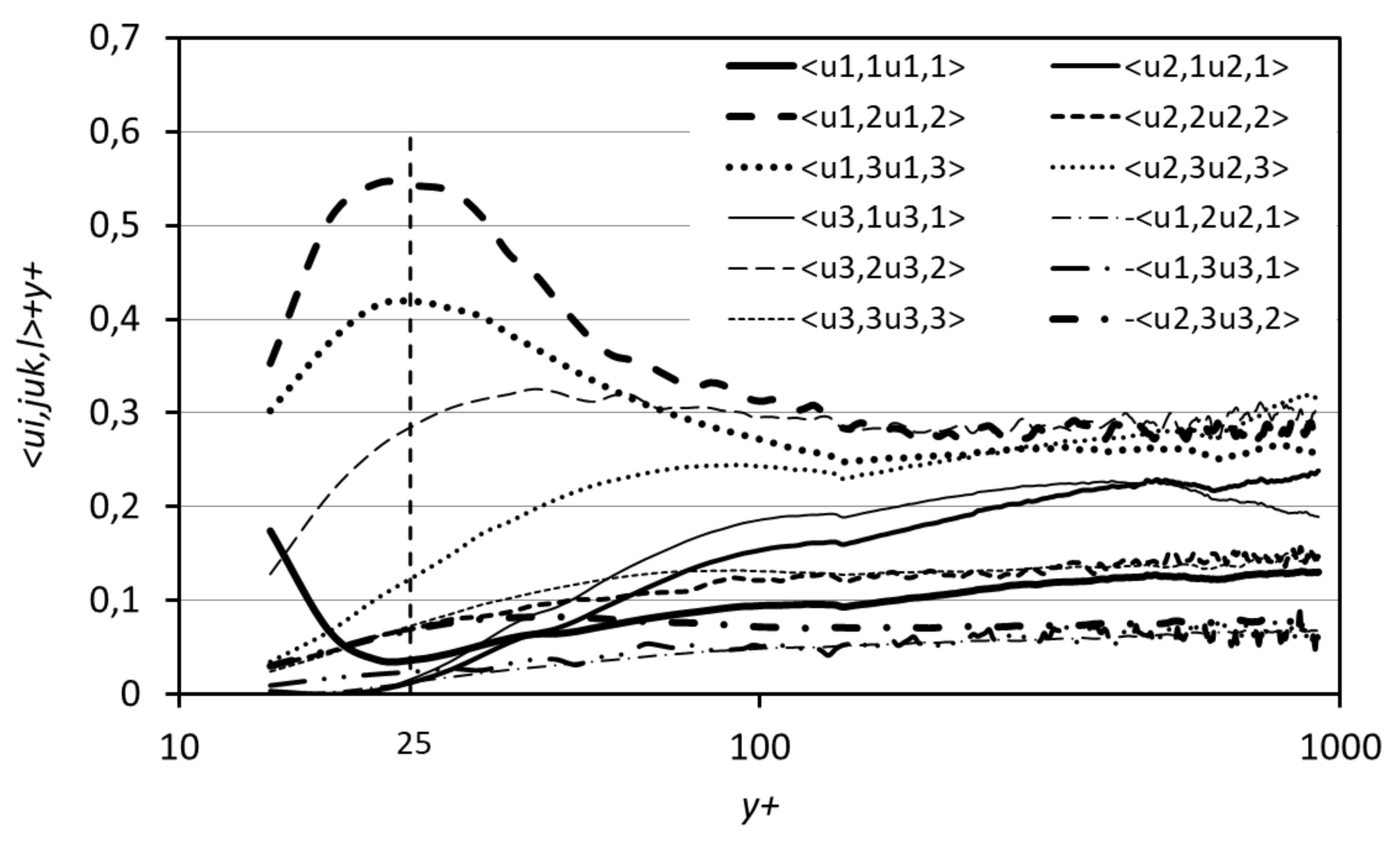}}
		\caption{Linear-log plot of all dissipation derivatives moments deduced from the PIV data times $y^+$ plotted together.}
		\label{fig:allderivativestimesy-lin-log}      
	\end{figure}
	
	In figure~\ref{fig:allderivativestimesy-lin-log}, all derivatives multiplied by $y^+$ are nearly constant, but it is impossible to tell whether the power is exactly $-1$ or simply close to it.  This is consistent with the arguments of \cite{george97b} that the overlap region of a boundary layers is slightly different from channel flow due to the streamwise inhomogeneity of the flow. For the mean velocity and turbulent velocity moments this implies weak power law profiles instead of logarithmic profiles.  Clearly the individual derivative moments presented in figures \ref{fig:allderivatives-lin-lin} to \ref{fig:allderivativestimesy-lin-log} show the same behavior.  As noted in Part III \cite{foucaut20}, decreasing spatial resolution as the wall is approached complicates a definitive answer from these data alone, as does the theoretical possibility of a non-zero offset, say $\varepsilon ^+ =  D/(y^+ + a^+)$~\cite{george97b,wosnik00,hultmark2012}.  Clearly we need much higher Reynolds number data to sort these theoretical nuances out properly.  But as Appendix~\ref{app:asymptotic_review} shows, simply performing experiments and multiplying data by y (and not testing multiple possibilities) will not be very useful.
	
	\subsection{Comparison with DNS and previous hot-wire results}
	
	\begin{figure}
		\resizebox{1.0\linewidth}{!}{\includegraphics[scale=1]{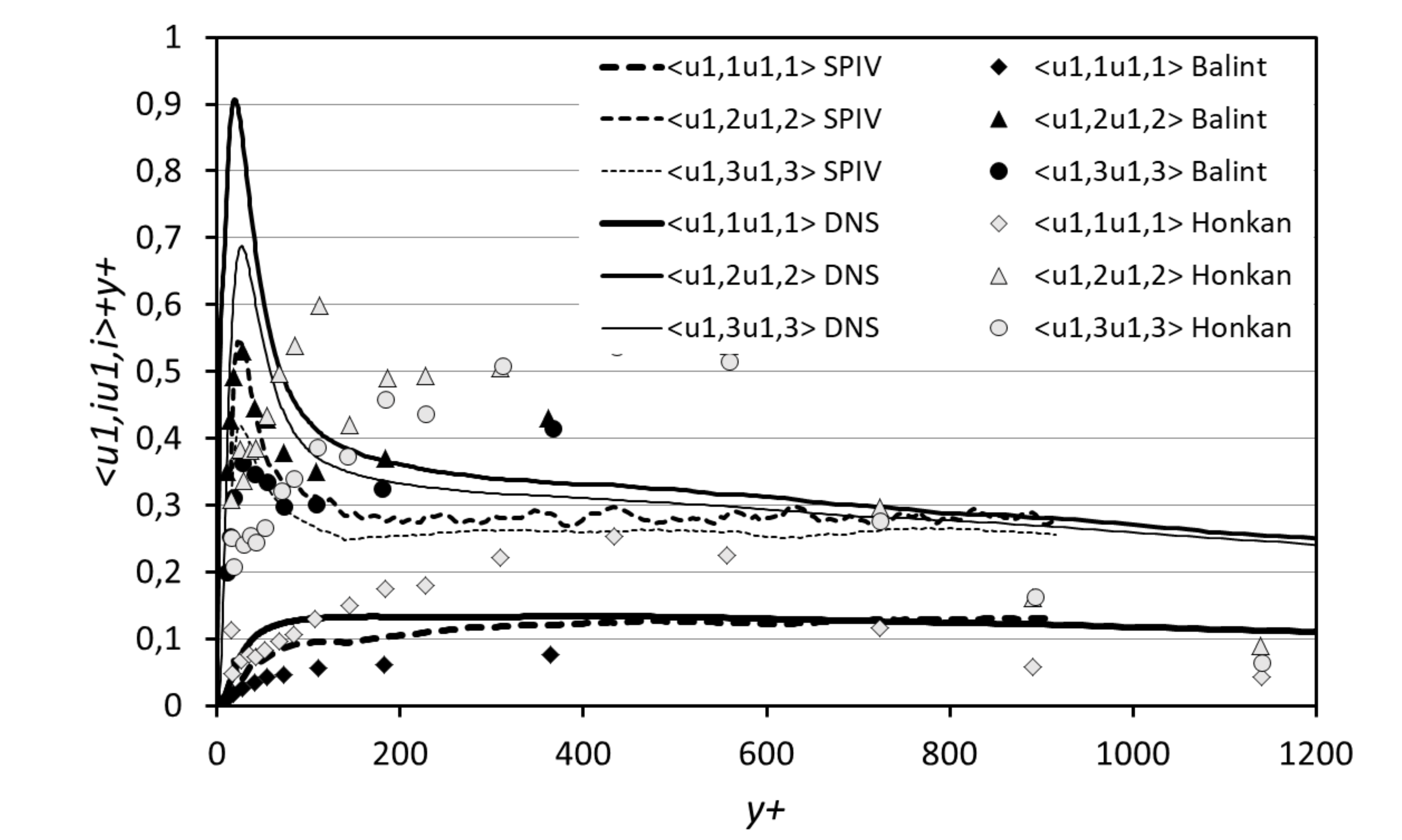}} \\
		\caption{Comparison of SPIV derivative moments involving $u_1$ with DNS and hot-wire results of \cite{balint91} and \cite{honkan97}.}
		\label{fig:terms_u1}      
	\end{figure}
	
	\begin{figure}
		\resizebox{1.0\linewidth}{!}{\includegraphics[scale=1]{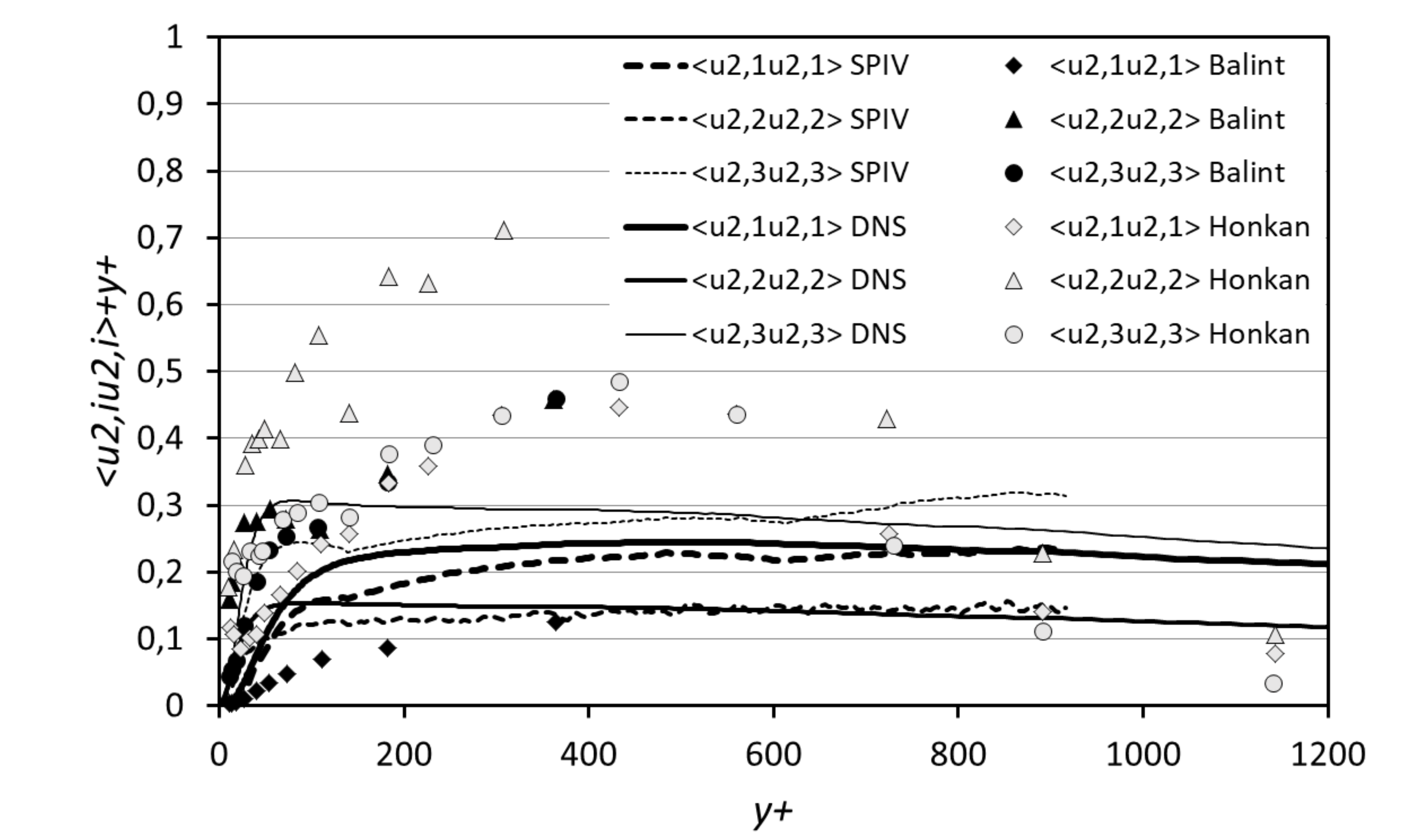}} \\
		\caption{Comparison of SPIV derivative moments involving $u_2$ with DNS and hot-wire results of \cite{balint91} and \cite{honkan97}.}
		\label{fig:terms_u2}      
	\end{figure}
	
	\begin{figure}
		\resizebox{1\linewidth}{!}{\includegraphics[scale=1]{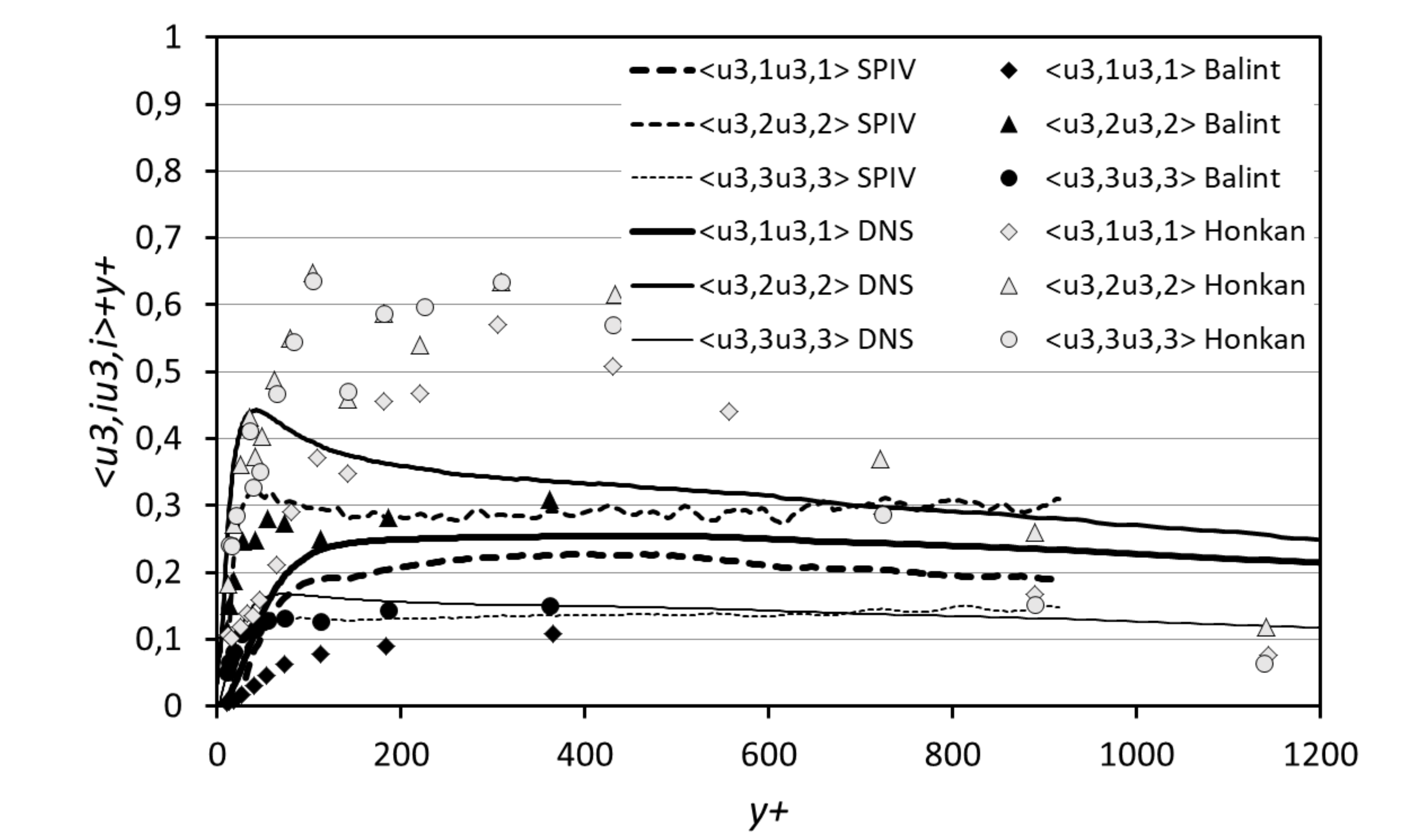}} 
		\caption{Comparison of SPIV derivative moments involving $u_3$ with channel DNS and hot-wire results of \cite{balint91} and \cite{honkan97}.}
		\label{fig:terms_u3}      
	\end{figure}
	
	\begin{figure}
		\resizebox{1.0\linewidth}{!}{\includegraphics[scale=1]{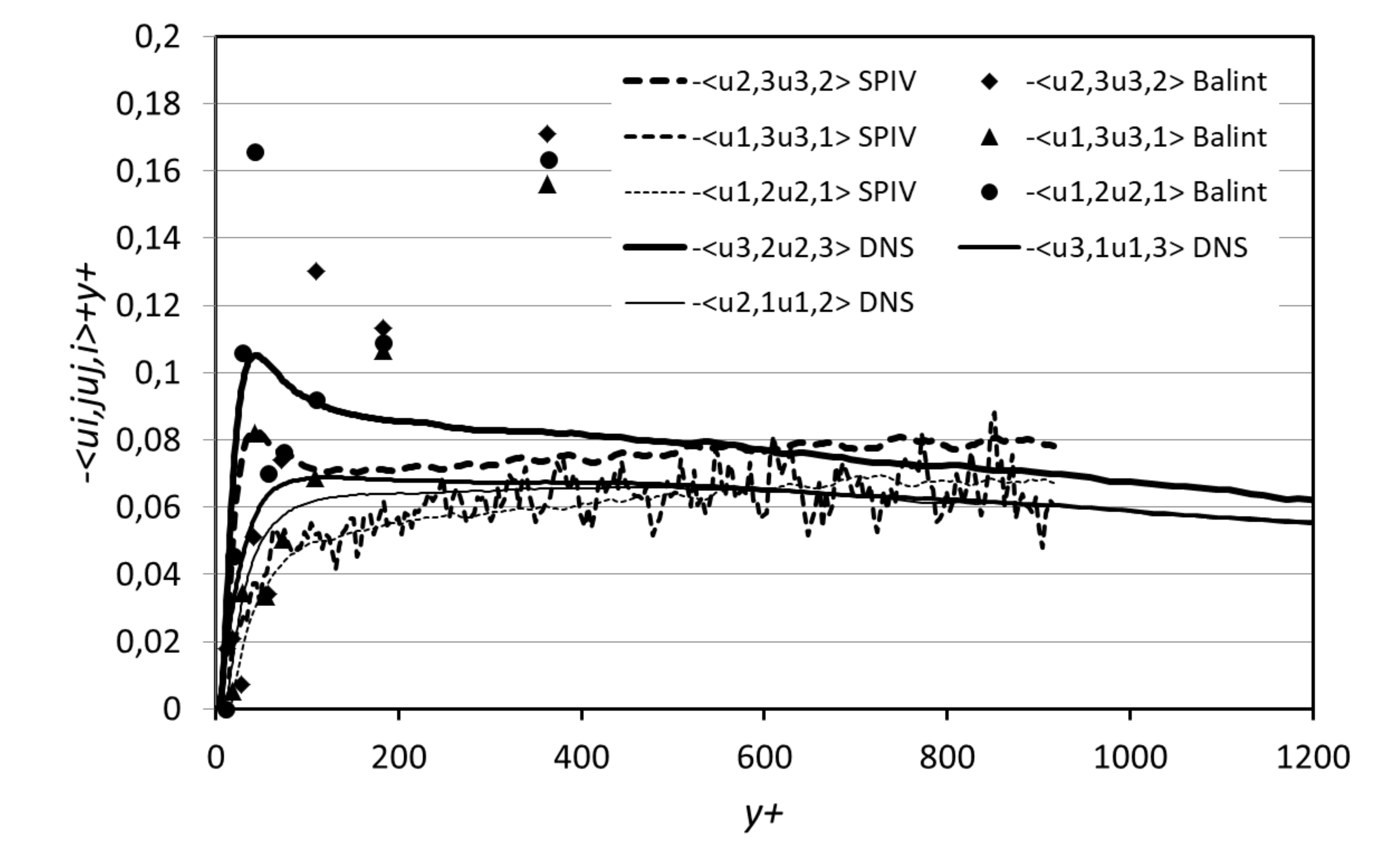}}
		\caption{Comparison of SPIV derivative cross-products with channel DNS and hot-wire results of \cite{balint91} and \cite{honkan97}.}
		\label{fig:terms_cross}      
	\end{figure}
	
	Figures~\ref{fig:terms_u1}  to \ref{fig:terms_cross} show normal and the crossed derivative moments, together with the channel flow DNS  described in section \ref{DNS-description} and previous hot-wire measurements of \cite{balint91} and \cite{honkan97}. All data are premultiplied by $y^+$ to evidence the differences. 
	Overall, despite a slight underestimation when approaching the wall, the agreement between the SPIV and the channel flow DNS is quite gratifying (at least in figures~\ref{fig:terms_u1} to \ref{fig:terms_u3}), despite the difference in outer flow. The result is as expected since the comparison is done mostly in the overlap and viscous regions. The situation is not the same for the previous measurements of \cite{balint91} and \cite{honkan97}. For the variances, the results of \cite{balint91} are in reasonable agreement with the DNS and the present data, except for $\langle u_{3,1} u_{3,1} \rangle$ which is very low. The crossed derivative moments are quite far off. This underestimation of the variances could be due to the difference in Reynolds numbers (which is not that large), but more likely is a consequence of limitations of hot-wire techniques. The data of \cite{honkan97} appear underestimated very near the wall (especially in figure~\ref{fig:terms_u1}) and significantly overestimated away from it. As will be seen, it appeared in the course of the preparation of this paper, that the $\varepsilon$ values in \cite{honkan97} were by error multiplied by 2 (this was confirmed by one of the authors). It is possible that the same error affects the individual variances (the crossed moments were not provided in the \cite{honkan97} paper). They would then come globally much closer to the other data. In any case, the hierarchy of the different variances is coherent with the other data in both figures~\ref{fig:terms_u1}, \ref{fig:terms_u2} and ~\ref{fig:terms_u3}. 
	Close to the wall the present SPIV results definitely underestimate the DNS derivative moments due to the spatial filtering of the smaller scales by the finite SPIV interrogation window (which is about $5\eta \times 5\eta$ in the present case). Farther away from the wall (outside $y^+ > 50$) the present results are in relatively good agreement with \cite{balint91} for some of the moments. It is difficult to find an explanation for these discrepancies as the technique to calculate these derivatives from the multiwire probes data is quite different from the SPIV approach. It is also different for the two HW data of \cite{balint91} and \cite{honkan97}, but involves in both cases a combination of first order spatial and time derivatives with a Taylor hypothesis to convert them in streamwise derivatives. In the SPIV, all derivatives are computed spatially on a uniform grid and using an optimized derivative filter.

	\subsection{Dissipation and production}
	
\begin{figure}
	\resizebox{0.95\linewidth}{!}{\includegraphics[scale=1]{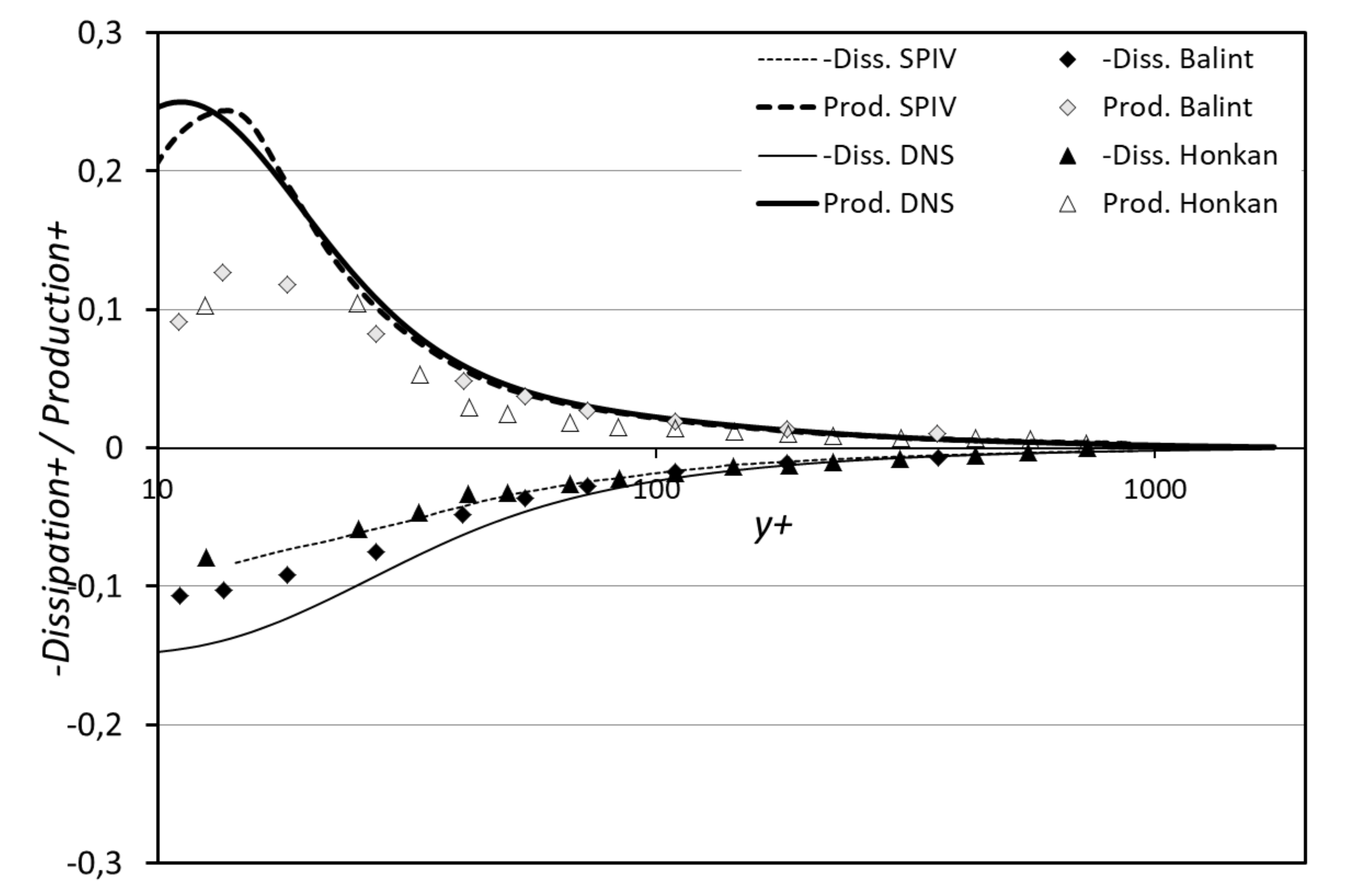}} 
	\caption{The dissipation rate $\epsilon$ and production rate in inner variables (linear-linear) for DNS and SPIV along with \cite{balint91,honkan2001}}.
	\label{fig:variousdissipproduction_linlin}     
\end{figure}

\begin{figure}
	\resizebox{0.95\linewidth}{!}{\includegraphics[scale=1]{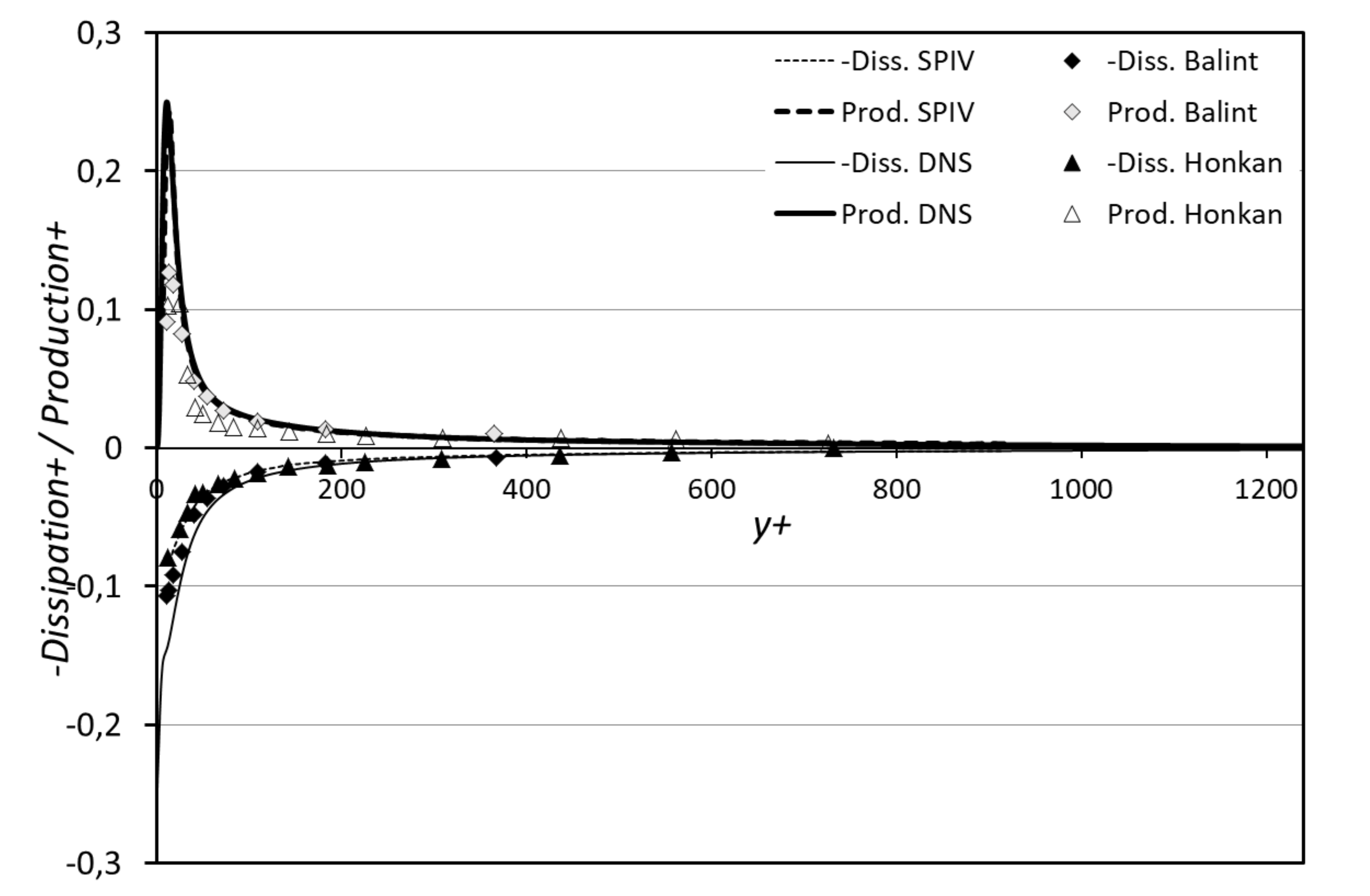}} 
	\caption{The dissipation rate $\epsilon$ and production rate in inner variables (linear-logarithmic) for DNS and SPIV along with \cite{balint91,honkan2001}.}
	\label{fig:variousdissipproduction_linlog}     
\end{figure}

\begin{figure}
	\resizebox{0.95\linewidth}{!}{\includegraphics[scale=1]{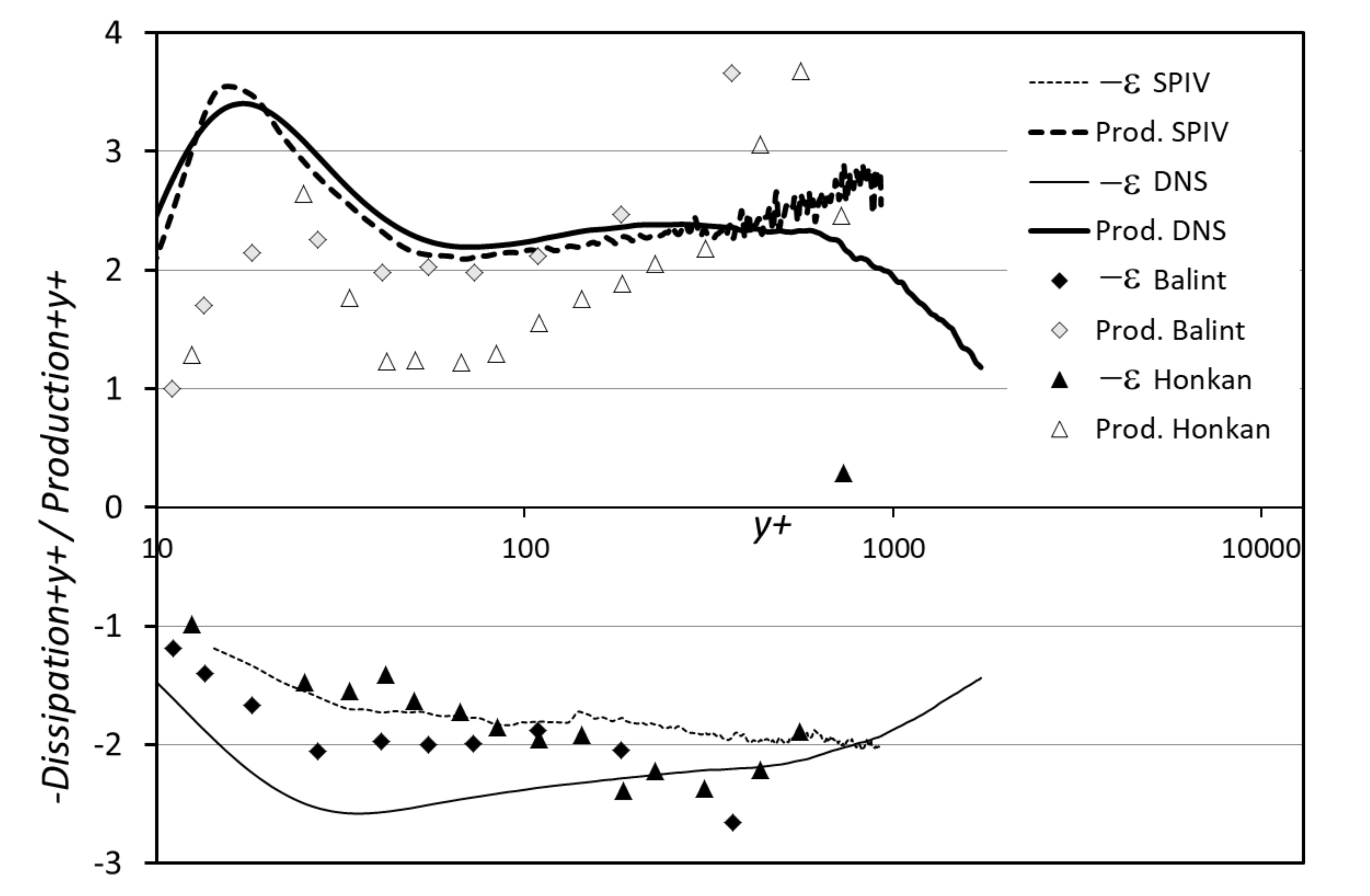}} 
	\caption{The dissipation rate $\epsilon$ and production rate in inner variables (linear-logarithmic) for DNS and SPIV along with \cite{balint91,honkan2001}. The data is pre-multiplied by $y^+$.}
	\label{fig:variousdissipproduction_premul}     
\end{figure}

	Figures~\ref{fig:variousdissipproduction_linlin}, \ref{fig:variousdissipproduction_linlog} and \ref{fig:variousdissipproduction_premul} show the dissipation and production rates from the SPIV and DNS, along with that of \cite{balint91} and  \cite{honkan97,honkan2001}. The data are presented successively in linear-linear, linear-logarithmic and linear logarithmic premultiplied by $y^+$. Outside of $y^+ \approx 50$, both the linear-linear and the linear-log plot of figures ~\ref{fig:variousdissipproduction_linlin} and \ref{fig:variousdissipproduction_linlog} suggest that all the results are in reasonable agreement. \footnote{Note the dissipation data of \cite{honkan2001} have been divided by 2
		since the data reported in the paper were in error. This is in agreement with a discussion with Prof. Andreopolis.} The pre-multiplied by $y^+$-plots of figure~ \ref{fig:variousdissipproduction_premul}, however,  tell a different story. The linear departures at about $y^+ \approx 100$ of the data of \cite{honkan97} can be accounted for by a constant additive noise, which would not be surprising given the hot-wire and data-processing noise and the decreasing signal level  with increasing distance from the wall.  By differentiating linear fits
	to these departures the noise level can be determined and subtracted. Figure \ref{fig:dissipationproduction_premul_cor} shows the  values of \cite{honkan97} corrected in this manner, and they are in much better agreement with the present data. The nearly horizontal data from
	about $100 \le y^+ \le 800$ of both the SPIV and the \cite{honkan97} data suggest strongly a near $1/y^+$ behavior in this region. This result is somewhat supported by the DNS data of \cite{sillero13} presented in figure \ref{fig:compare_dnsdissipation}. This will be discussed in detail below; but it is already interesting to note that in the same region, the DNS data for $\varepsilon$ seem to depart slightly from this $1/y^+$ behavior while the production is closer. Note that no virtual origin as been used in the premultiplied plots, but this plot can be made significantly flatter by choosing one equal to about 7.  
	
	\begin{figure}
		\resizebox{0.95\linewidth}{!}{\includegraphics[scale=1]{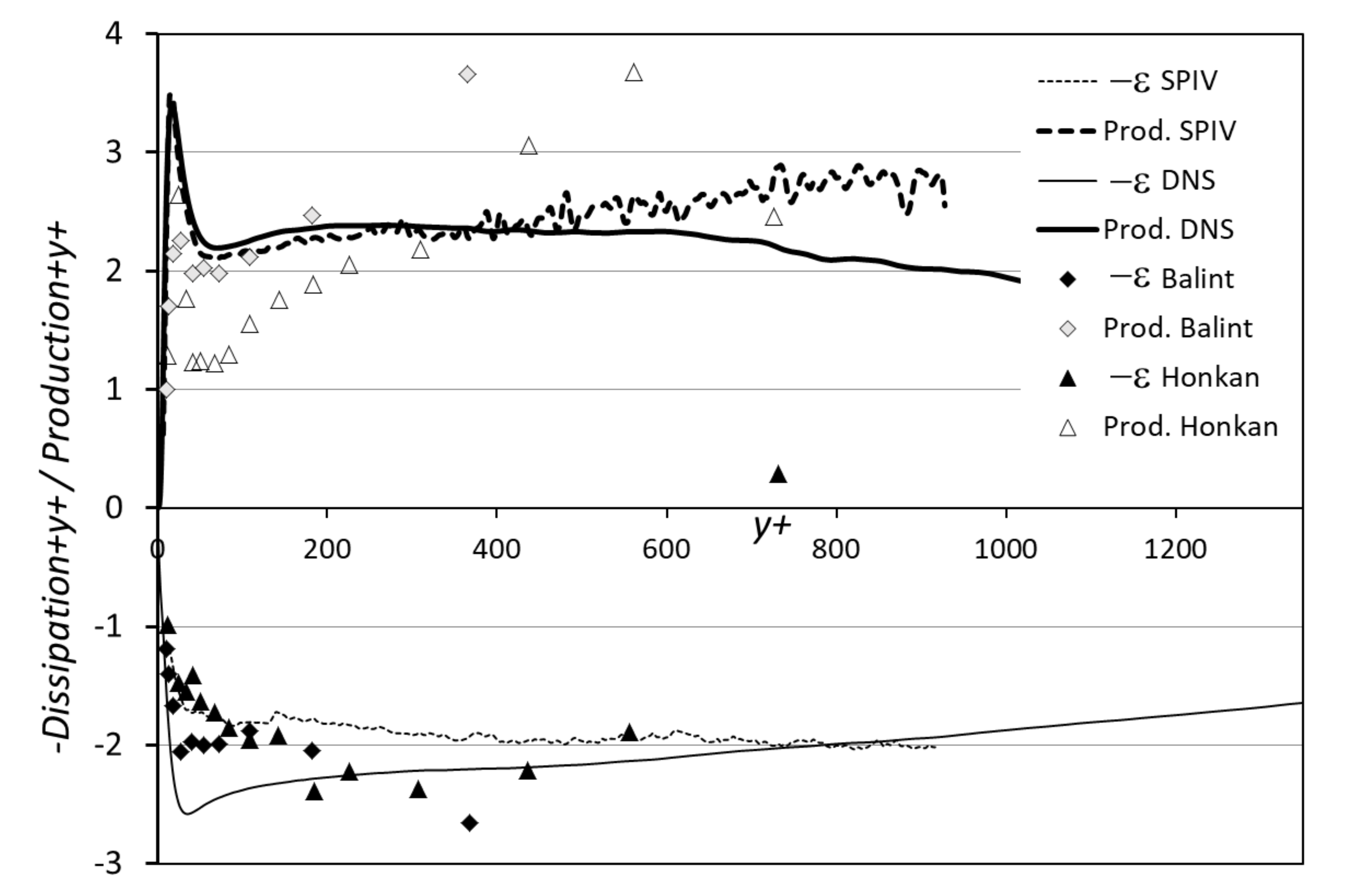}} 
		\caption{The dissipation rate $\epsilon$ and production rate in inner variables (linear-logarithmic) for DNS and SPIV along with \cite{balint91,honkan2001}. The data is pre-multiplied by $y^+$.}
		\label{fig:dissipationproduction_premul_cor}     
	\end{figure}
	
	It is commonly assumed that the near wall peaks observed in the linear-log plots of figure \ref{fig:variousdissipproduction_linlog} (or even more visible in log-log plots) mean that the near wall region dominates the overall dissipation for the boundary layer. The opposite is in fact in evidence from the linear-linear plot of Fig.~\ref{fig:variousdissipproduction_linlin}, where the long tail with increasing distance from the wall is the primary contributor to the integrated dissipation across the boundary layer.  In fact since the near wall peak is fixed in inner variables, $y^+ = y u_\tau/\nu$, while the tail is in outer variables, the relative contribution to the integral shifts progressively away from the wall as the Reynolds number increases (v.\ \cite{george97b}). This will be more obvious when comparing measurements from different Reynolds numbers; in particular, the outer variables plots where the ordinate is $\bar{y} =y/\delta_{99}$.  In these outer variables, most of the profile is unchanged with increasing the Reynolds number except for the near wall region which is universal in $y^+$ variables, and moves progressively toward smaller values of $\bar{y}$ when $\delta_{99}^+$ increases.  Thus at high Reynolds number the dissipation contribution to the boundary layer is effectively just an integral under the `tail', most of which is in the overlap (or `log') region.\footnote{Note that the preferred term is `overlap' region, since it may not be a `log-region', but a power-law. This is nearly impossible to determine from experimental data alone \cite{george97b,george2007}.}

	\subsection{Mean square vorticity}
	
	\begin{figure}
		\resizebox{0.95\linewidth}{!}{\includegraphics[scale=1]{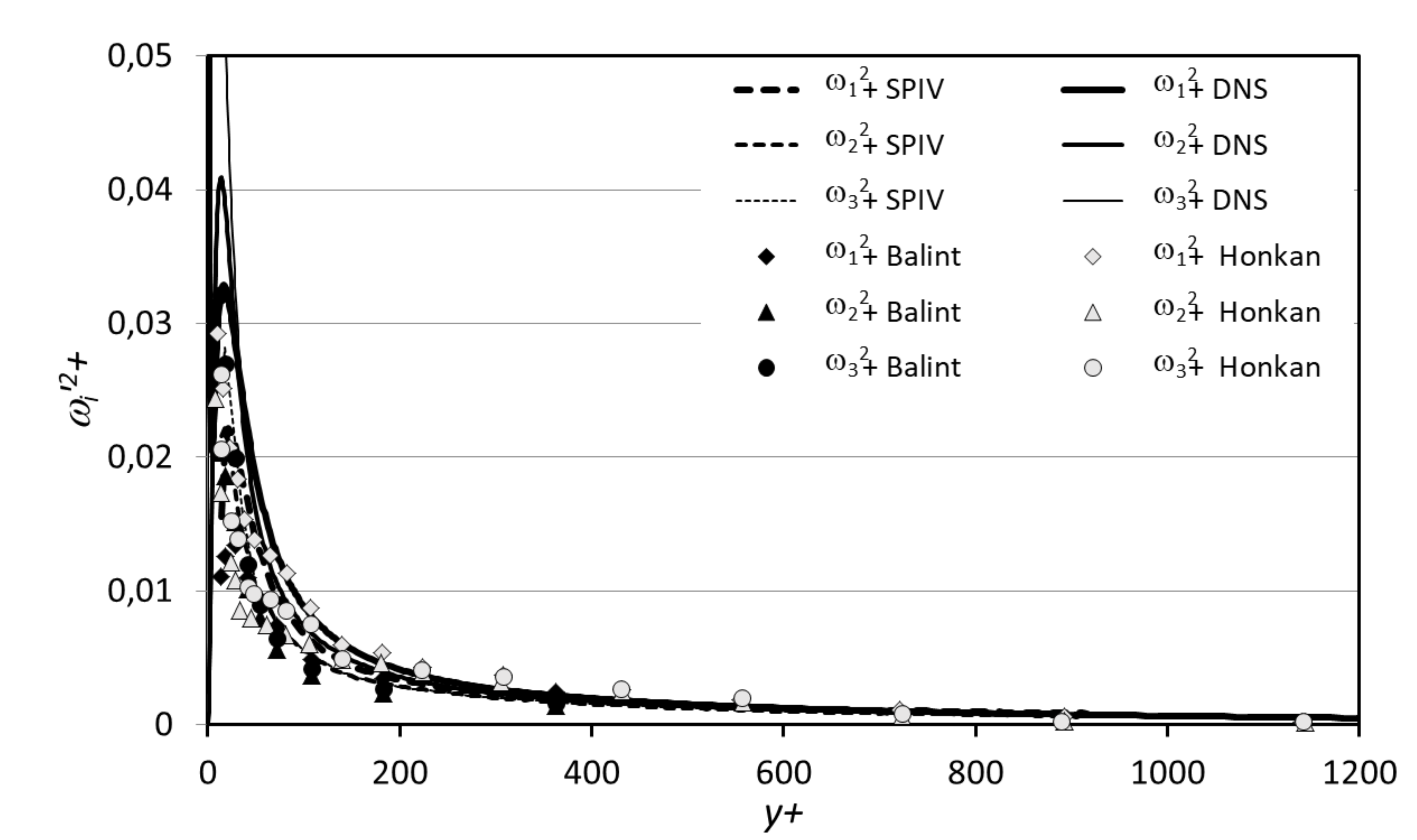} }
		\caption{Mean square vorticity (enstrophy) components scaled in wall units (lin-lin plot).}
		\label{fig:variousvorticity}      
	\end{figure}
	
	\begin{figure}
		\centering 
		\includegraphics[width=0.99\columnwidth]{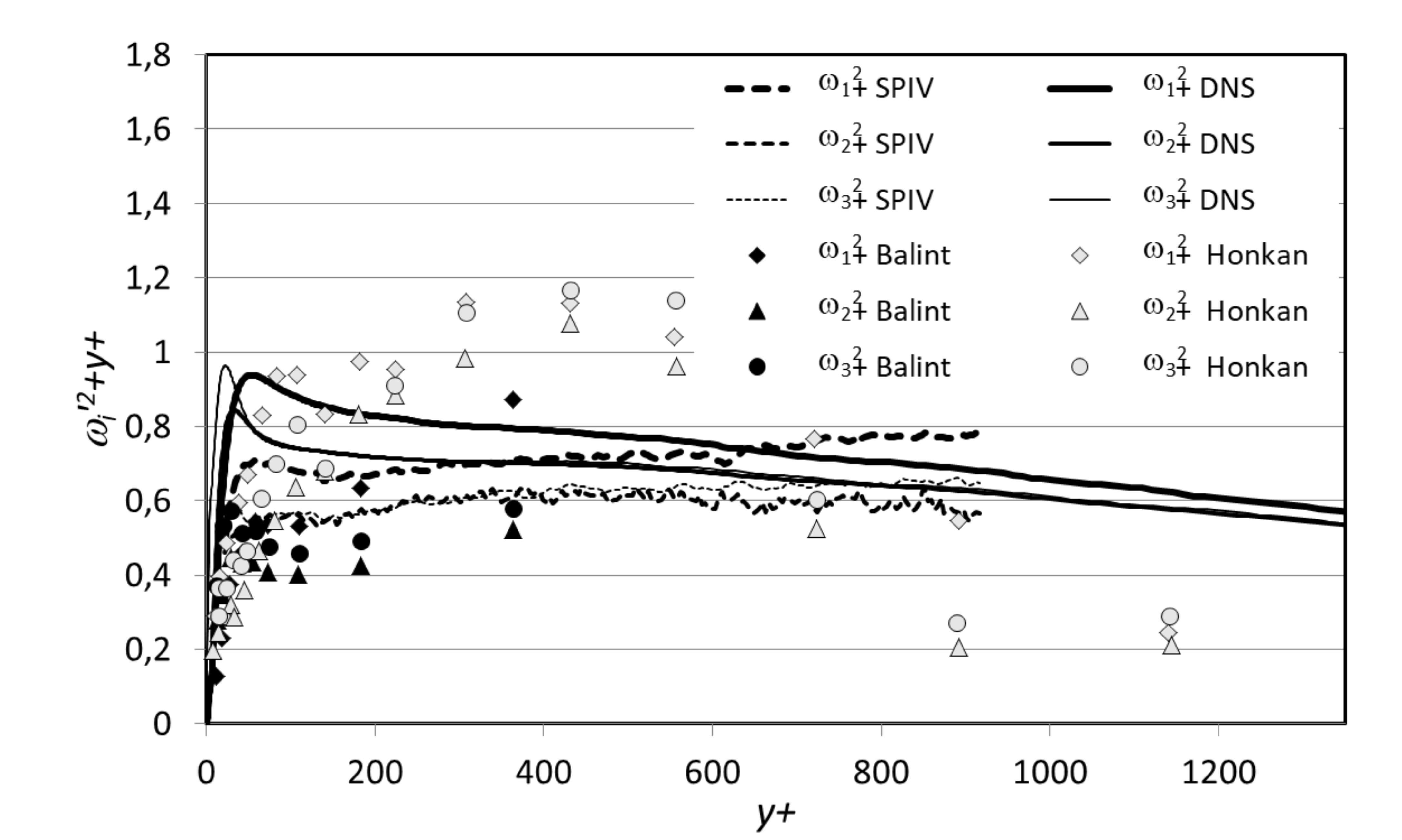}
		\caption{Mean square vorticity (enstrophy) components scaled in wall units and multiplied by $y^+$ (lin-lin plot). }
		\label{fig:vorticity_premulti}
	\end{figure}

	Linear-linear plots of the mean square vorticity (enstrophy) from the same data sets are given in Figure~\ref{fig:variousvorticity}. It is also provided with the vertical axis multiplied by $y^+$ in figure \ref{fig:vorticity_premulti}. In the first representation, the global agreement looks fairly good in the outer part, say above $y^+ = 100$. In particular, the data of \cite{honkan97} are quite good. It is much more difficult to come to some conclusion near the wall. In the premultiplied plot, substantial differences can be observed between the different data sets. The data of \cite{balint91} are quite low but show the right tendency, although they do not go far enough from the wall to draw a definite conclusion. The data of \cite{honkan97} show a strange behaviour. As far as the SPIV and DNS are concerned, although these are different flows, the 20\% difference between them suggests that the spatial resolution of the SPIV data may be worse than our estimates, at least for the moments needed to compute the enstrophy. Here again, the different enstrophy components from the SPIV show a fairly good $1/y$ behaviour in the overlap region. Again, the DNS is showing a weak slope which could be compensated by a shift of origin.

\section{Implications for understanding the overlap region \label{sec-asymptotictheory}}

There has been considerable revision in our thinking over the past few decades about where the overlap region is located and exactly what it is.  The whole idea is based on proper scaling of the near wall and outer regions of the flow, then working out how the pieces fit together.  The experimental results provided herein make it clear that at least the predictions for the dissipation in the overlap region appear to have validity.  But is there more?  

Appendix \ref{app:asymptotic_review} provides a brief historical review of the asymptotic methodology used.  This section explores two new implications of that methoodology:  first the behavior of the transport moments; and second, the implications for most turbulence models.
To this point there has to the best our knowledge been no discussion or tests of these ideas. 

\subsection{The non-zero transport term in the overlap region \label{sec-transport}}

\begin{figure}
	\centering 
	\includegraphics[width=0.99\columnwidth]{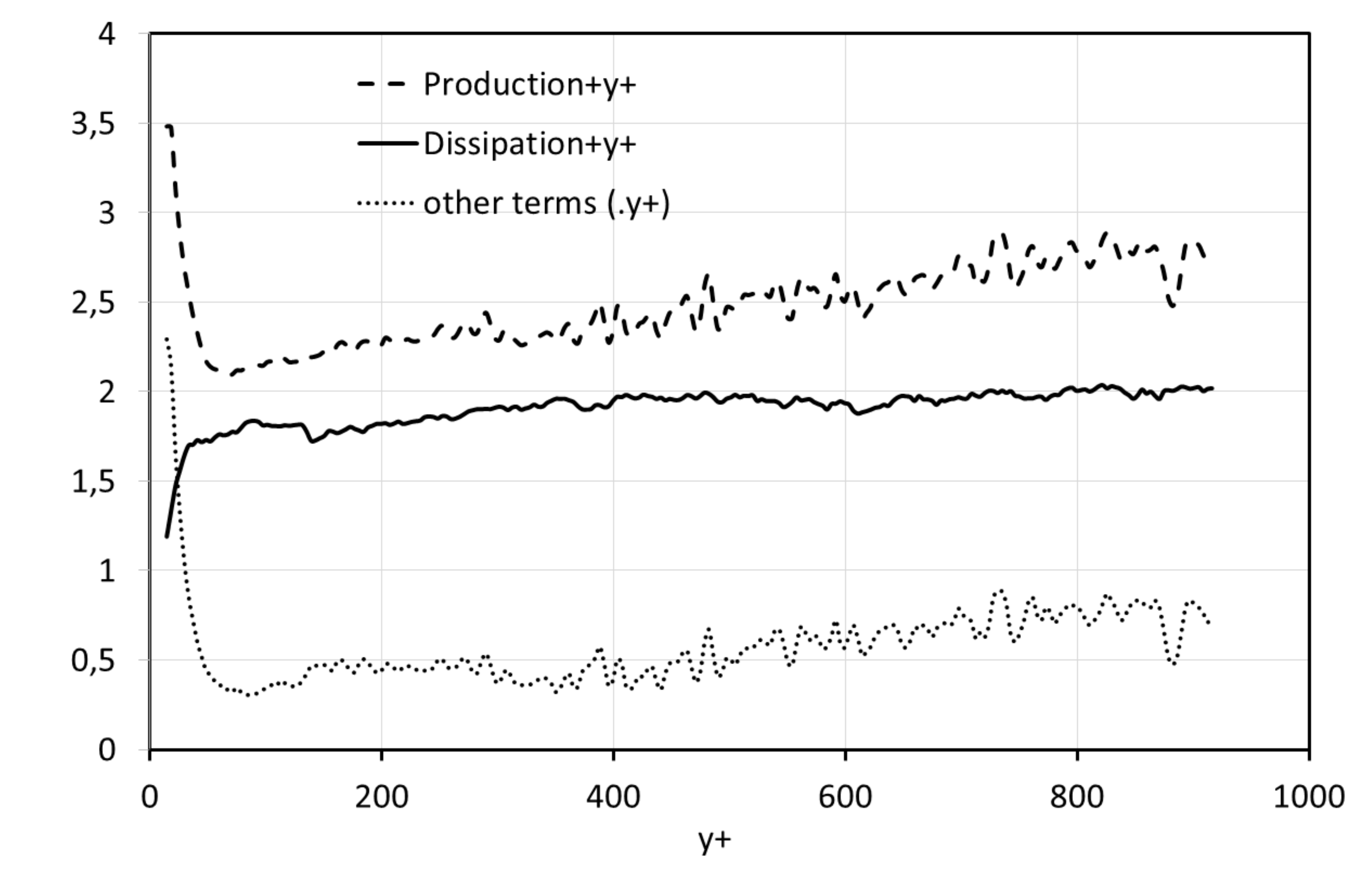}
	\caption{Linear plot of the SPIV energy balance times $y$ in inner variables; i.e.,  production, $y^+\langle uv \rangle^+ dU^+/dy^+$ (broken), dissipation $y^+ \varepsilon^+$ (plain)  and the other terms $d[-\langle \rho u_iu_i u_j \rangle - \langle p u_j \rangle + \nu \langle s_ij u_i \rangle]^+ /dy^+$ (doted). }
	\label{fig:dissipation_premulti_lin}
\end{figure}

Figure~\ref{fig:dissipation_premulti_lin} shows a plot of the production and dissipation times $y^+$, along with a plot of the difference $\mathcal{P}^+ - \varepsilon^+$ as obtained from the boundary layer SPIV data. Clearly, in the overlap region, there is about a 10\% difference.  Balint et al.~\cite{balint91} also observed the same thing.  They further noted that the suggestion of Tennekes and Lumley~\cite{TL72} that production and dissipation should be equal in the overlap region seemed to be only approximately satisfied in their measurements. The discrepancy is too consistent to be measurement error, and has approximately a $1/y^+$-dependence as does the production in
this region. The only other term in the energy balance that could be significant in this region is the turbulence transport term. The slight deviation from horizontal of the dissipation  and the difference are most likely due to the need for a virtual origin to account for the mesolayer ~\cite{george97b,wosnik00}.  Or, as noted above, that they should be described by a power-law slightly different from $-1$.  Experiments at much higher Reynolds number should be able to sort this out.

\begin{figure}
	\centering 
	\includegraphics[width=0.99\columnwidth]{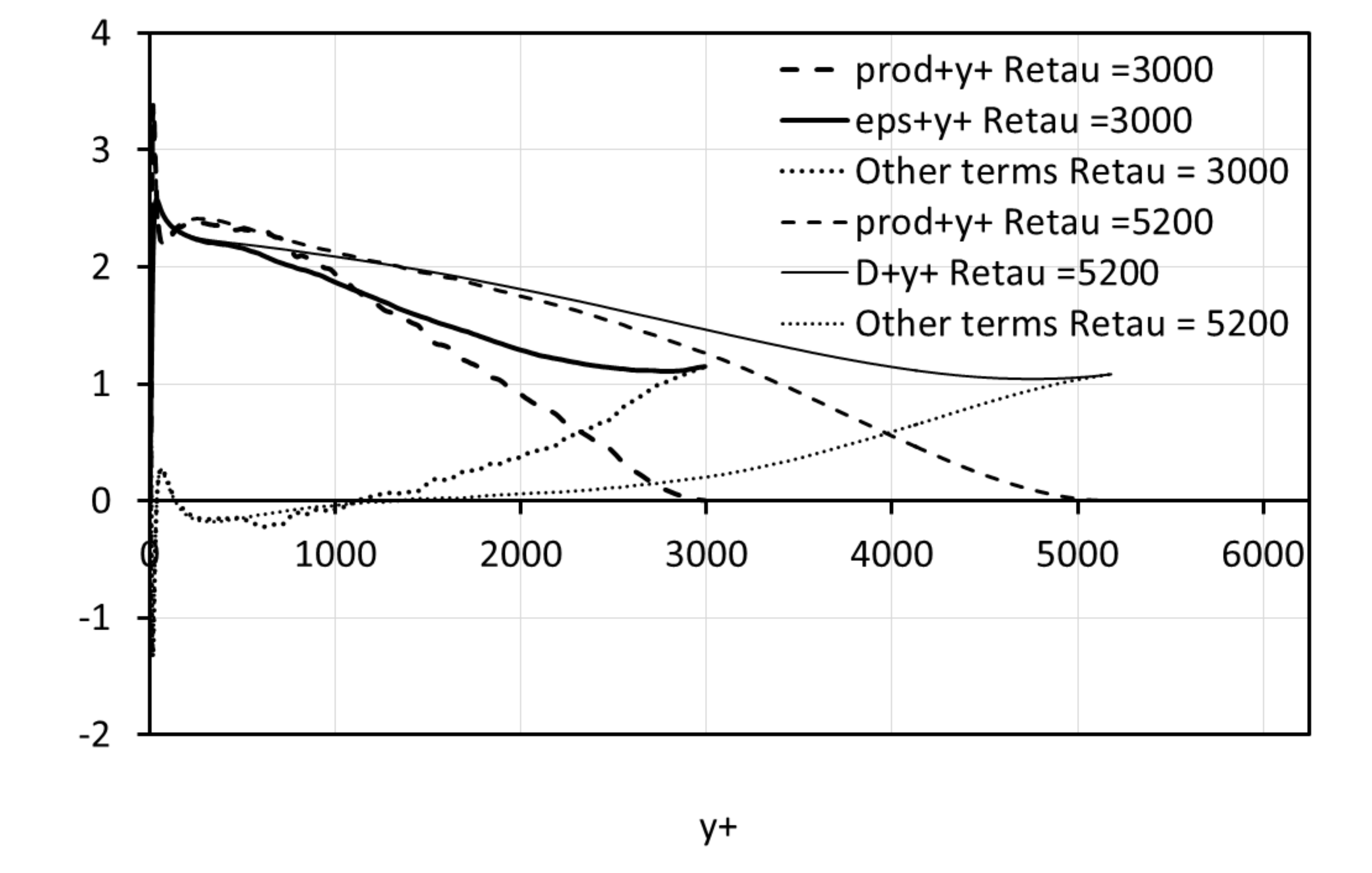}
	\caption{Linear plot of the DNS energy balance times $y$ in inner variables; i.e., production, $\langle uv \rangle^+ dU^+/dy^+$ (broken), dissipation $y^+ \varepsilon^+$ (plain)  and the other terms $d[-\langle \rho u_iu_i u_j \rangle - \langle p u_j \rangle + \nu \langle s_ij u_i \rangle]^+ /dy^+$ (doted), for two values of the Reynolds number. }
	\label{fig:compare_dnsdissipation_lin}
\end{figure}

\begin{figure}
	\centering 
	\includegraphics[width=0.99\columnwidth]{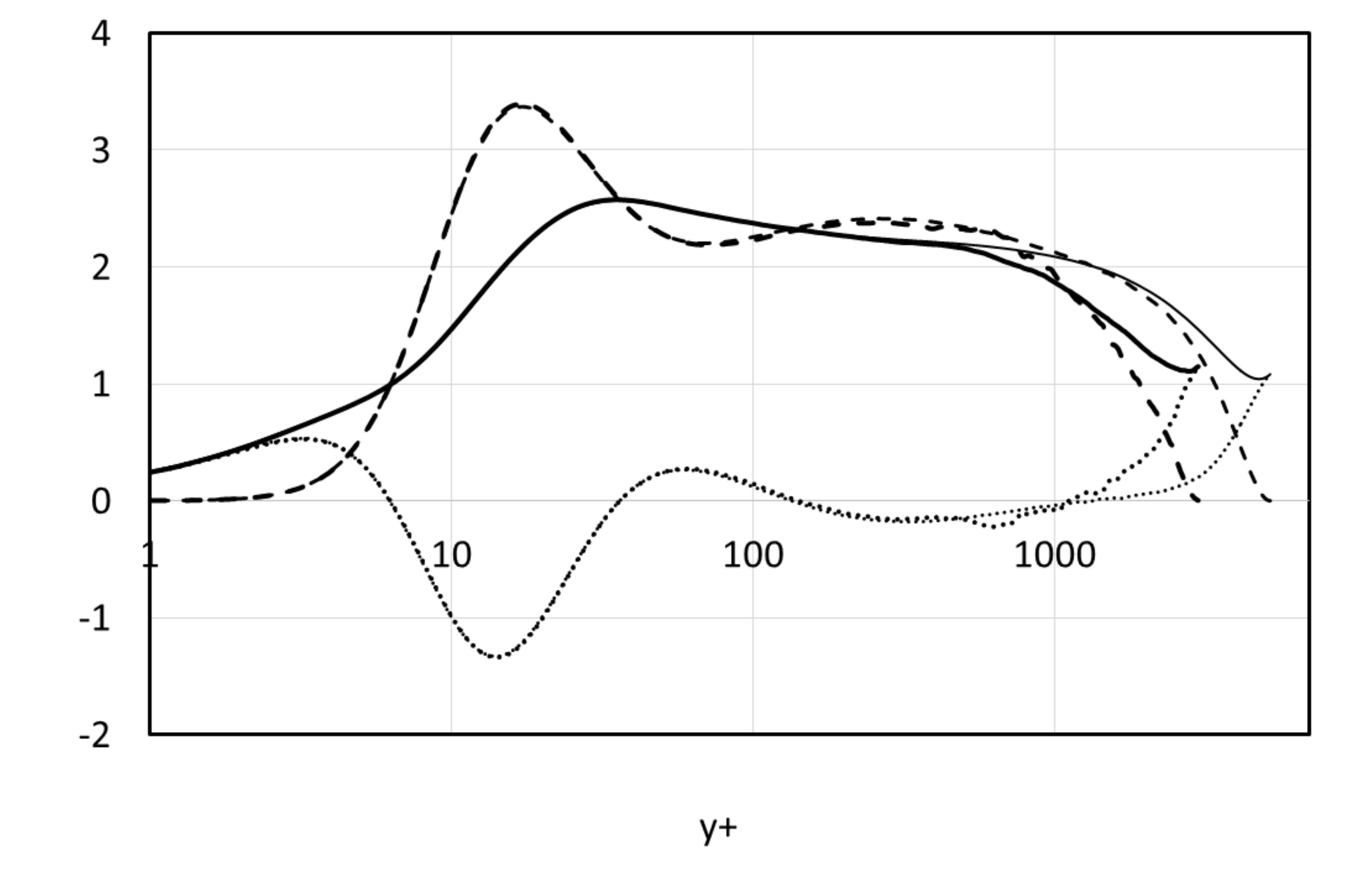}
	\caption{Logarithmic plot of the same data as in figure \ref{fig:compare_dnsdissipation_lin}.}
	\label{fig:compare_dnsdissipation_log}
\end{figure}

Figures~\ref{fig:compare_dnsdissipation_lin} and \ref{fig:compare_dnsdissipation_log} show a similar plot for the channel DNS in respectively linear and logarithmic representation. Two sets of data are provided: the present DNS data of \cite{thais11} at $Re_{\tau} = 3000$ and the DNS of \cite{lee15} at $Re_{\tau} = 3000$.  Up to $y^+ = 600$, the general features are comparable to the boundary layer experimental data, and again there is about a 10\% discrepancy. But here the difference is definitely due to transport terms in the energy balance equation. Above $y^+ = 600$, the behavior of the channel data changes significantly. It is difficult to draw some definite conclusions as the boundary layer data extend only up to $y^+ = 1000$, but they do not show any tendency to decrease yet. If confirmed further away from the wall, this is most likely evidence of a different physics between the boundary layer and channel in the outerpart as seems to be supported by the DNS data of figure \ref{fig:compare_dnsdissipation}. It is of interest to note the effect of the Reynolds number between the two DNS which show a very similar behaviour and definitely support the idea of a nearly $1/y$ behaviour of both production and dissipation in the overlap region when the Reynolds number is high enough. This idea is also supported by \cite{smits11} in their figure 1.

It is easy to extend the same Near-Asymptotic analysis to these clearly important transport terms.  For example, for a streamwise homogeneous flow, they can be shown to be logarithmic; i.e.,
\begin{equation}
	\langle -\frac{1}{2} u_iu_i u_j \rangle^+ - \frac{1}{\rho}\langle p u_j \rangle^+ +2 \nu \langle u_i s_{ij} \rangle ^+ = T ~ln y^+ + S.
\end{equation}
where $T$ and $S$ are related to the logarithmic profile counterparts of the velocity profile.
As a result,  the derivative in the wall normal direction is
$1/y^+$, exactly as observed in the figure~\ref{fig:compare_dnsdissipation}.

For the boundary layer, these same transport terms vary as a weak power law for a streamwise developing flow; i.e,

\begin{equation}
	-\langle \frac{1}{2} u_iu_i u_j\rangle^+ - \langle\frac{1}{\rho}p u_j \rangle^+ + 2 \nu \langle u_i s_{ij} \rangle^+ = T_i  {y^+}^\beta
\end{equation} 

The implications of this are most easily illustrated by the example below.
In the overlap region of a channel flow the mean convection terms of equation~\ref{eq:k} for the turbulent kinetic energy are exactly zero and the viscous terms are negligible, so the overall energy balance reduces to just

\begin{equation}
	0  = \frac{d}{dy^+}  \left\{-\frac{1}{2} \langle u_iu_iu_j \rangle^+ - \frac{1}{\rho}\langle p u_j \rangle^+ + 2 \nu \langle u_i s_{ij} \rangle^+ \right\}
	-\langle uv \rangle^+ \frac{dU^+}{dy^+}  - \varepsilon^+
	\label{eq:overlap_dissipation}
\end{equation}
But in inner variables (i.e., normalized by $u_\tau$ and $\nu$), $-\langle u v \rangle^+ = 1$ and $dU^+/dy^+ = 1/\kappa$ (where $\kappa$ is the Von-Karman parameter), so it follows immediately that ALL the surviving terms in the overlap region vary as $1/y^+$, and  $0 = T + (1/\kappa) - D$.  Similar relations apply for the power law relations in the developing boundary layers. Clearly these can have the same $y^+$-dependence ONLY if $\beta =\gamma$, the same power as the mean velocity.   And the equations can balance only if  $0 = T_i + B_i -D_i$.

Like all previous measurements it is impossible to distinguish between log or power, so  it remains for subsequent experiments to sort this out.  Note that the important test will not be curve-fits alone, but which outer scaling law works the best, $u_\tau^4/\nu$ or $u_\tau^2 U_\infty/\delta$.

\subsection{An implication for turbulence modeling \label{sec-eddyviscosity}}

The results of the preceeding section provide a direct way to evaluate one of the fundamental {\it empirical} tenets of turbulence modeling; namely the idea of using gradient transport to model higher order moments. This will not be a detailed presentation of turbulence models, but hopefully it will provide some clues for those who seek to improve them.

The ideas of asymptotic analysis are based on fundamental ideas that can be related directly to similarity ideas for the equations of motion -- at least when implemented using the equilibrium similarity ideas of \cite{george97b}.  So any turbulence model should  be consistent with it.  It is easy to show that the usual eddy viscosity is not.
Typically an eddy (or turbulence) viscosity
is chosen like this:
\begin{equation}
	\nu_t \propto k^2/\varepsilon.
\end{equation}
This idea is basically using K41 \cite{kolmogorov41} to argue for a turbulence length and velocity scale. For simplicity we consider only the streamwise homogeneous theoretical results.  

As noted in Appendix~\ref{app:asymptotic_review} and \cite{hultmark2012}, $k$ is logarithmic; say $k =  A \ln y^+ + F $.  And as we have seen herein, $\varepsilon ^+$ varies as $D/y ^+$.  So 
\begin{equation}
	\nu_t^+ \propto  [A \ln y^+ + F]^2 [y^+/D]
\end{equation}
This would imply that the transport terms vary as

\begin{equation}
	-\langle \frac{1}{2} u_iu_i u_j - \frac{1}{\rho}p u_j \rangle \propto \nu_t \frac{\partial k}{\partial y} \propto  [A \ln y^+ + F]^2,
\end{equation}
which differs by a factor of $ [A \ln y^+ + F]$ from the asymptotic behavior deduced in the preceeding section.

The problem of course is related to the above comment about the applicability of K41 assumptions to the overlap region of wall-bounded flows.  It simply assumes a behavior which cannot be corrected for by the usual near-wall correction terms; i.e., by adding a viscous term.  A possible  choice for this overlap region which is asymptotically correct would be to replace one of the $k$'s by the Reynolds shear stress, $-\langle uv \rangle$ since in this region $-\langle uv \rangle = u_\tau^2$; i.e.,
\begin{equation}
	\nu_t \propto \frac{-\langle uv \rangle  k }{\varepsilon}
\end{equation}
Note that this particular model gives the correct asymptotic result for both power law and log law theories. There are most likely other alternatives which may be better, but they should always be consistent with the asymptotic theory in the overlap region.

\section{Summary and Conclusions}

 For both turbulence modeling and theoretical developments it is of prime importance to characterize the dissipation rate. This key parameter is difficult to access in practice as it involves all the terms of the strain-rate tensor and measurement requires a very good spatial resolution. With this objective, a specific SPIV experiment was carried out in the turbulent boundary layer which allowed derivative  measurements of all three velocity components in all three space directions. Also, as a matter of validation and comparison, the data of a specific channel flow DNS at the same Reynolds number were processed to provide the same information.
 
 This first paper shows the measured dissipation in the boundary layer up to $ y^+ \le 1000$ ($y/\delta_{99} < 0.1$), and compares it to previous measurements and the channel flow DNS at similar Reynolds number.  All of the experimental results show some evidence of spatial filtering, but taken together with the DNS give considerable insight into the dissipation. 
 
For the first time, in the present experiment, all 12 terms contributing to the turbulent energy dissipation were measured directly using only spatial derivatives, and without any additional hypotheses. The results show that these terms behave quite differently in the inner part of the boundary layer. Despite an obvious spatial filtering, the hierarchy between the terms is in very good agreement with the DNS, except that the boundary layer data show generally a better $1/y$ behaviour in the overlap region than the channel flow DNS. This could be attributed to the difference in flow type, but needs further investigation. Both the SPIV and the DNS data confirm that all the derivative cross-products shown in figure \ref{fig:terms_cross} are negative. Compared to the variances given in figures \ref{fig:terms_u1}, \ref{fig:terms_u2} and \ref{fig:terms_u3}, they are about an order of magnitude smaller. They will be adressed in detail in part II \cite{george20}, together with the different classical hypotheses used to simplify $\varepsilon$ And a new theory for their behavior in boundary layer flows will be proposed.
 
 As far as this full dissipation is concerned, a thorough comparison with the data of \cite{honkan97} showed that, after dividing their $\varepsilon$ by 2 (in agreement with the authors) and removing the noise, they come in relatively good agreement with the present data and those of \cite{balint91}. The three experiments were run at different Reynolds numbers (the present one being the highest) but at very comparable spatial resolutions when compared to the Kolmogorov scales. So, apart from the noise, the differences between them can only be attributed to the difference in Reynolds number or in the method used to assess the derivative moments. The difference in Reynolds number is too small to account for the differences. So it is most likely the data processing which is the most reliable explanation. This is supported by the fact that in SPIV the spatial derivatives are all calculated directly, as the 3D velocity fields are provided in the three directions of space (while with the hot wires a Taylor hypothesis has to be used). As detailed in part III \cite{foucaut20}a very careful procedure was used to denoise and validate the SPIV data. Finally, these SPIV data are very consistent with the DNS results.  
 
 Another important result of the present contribution is that the theoretical behavior (\cite{george97b,wosnik00}) of the dissipation in the inertial layer as $\varepsilon^+ \approx D {y^+}^{-1}$ (or weak power law behavior) is confirmed for the boundary layer with $D \approx 2.0$. For the channel flow DNS, this is at least approximately true for $30 \le y^+ \le 600$, but a clear departure is observed in the outer region beyond $y/H = 0.2$, consistent with the very different physics in the outer flows.  
 
 The $1/y$ behaviour of the dissipation evidenced here allowed us to revisit the standard modelling of the diffusion terms by a  gradient hypothesis and to show that the present model of the eddy viscosity based on $k^2/\varepsilon$ leads to a theoretical inconsistancy in the overlap region. A model consistent with asymptotic analysis is proposed. It is surely not the only possible option and it needs to be validated by further direct measurement and DNS.
 Clearly further attention to the turbulence models is warranted as well. 

Also, as will be discussed in detail in part II \cite{george20}, $\varepsilon^+$ and $D^+$ are nearly indistinguishable over this range. This is quite surprising, since the hypothesis of local homogeneity near the wall breaks down inside $y^+ = 30$.   As noted above, an alternative hypothesis consistent with the data will be proposed in part II \cite{george20}, which also examines in detail the statistical characteristics of the dissipation and the classical hypotheses which are usually proposed to simplify the modelling approach: local isotropy, local homogeneity and local axisymmetry. The present data will clearly identify which of these hypotheses is valid.  
 Part III \cite{foucaut20}, which is a third paper, has been reserved to detail the experimental procedures and methodology, including an error analysis. Given the significant effort put into this, it was considered that a separate paper going into all the details would be helpful to the experimentalist community.

\section*{Acknowledgement}
This work was supported through the International Campus on Safety and Inter modality in Transportation (CISIT). This work was carried out within the framework of the CNRS Research Federation on Ground Transports and Mobility, in articulation with the Elsat2020 project supported by the European Community, the French Ministry of Higher Education and Research, the Hauts de France Regional Council. Centrale Lille is acknowledged for providing regular financial support to the visits of Pr. George and the "R\'{e}gion Nord Pas-de-Calais" for supporting his visit in 2009. This research has granted access to the HPC resources  of [CCRT /CINES /IDRIS] under the allocation i20142b022277 and i20162a01741 made by GENCI (Grand Equipement National de Calcul Intensif). L. Thais is  acknowledged for providing the data of his DNS of channel flow. 

\bibliographystyle{plainnat}
\bibliography{biblio_jfmdissipation}

\appendix

\section{Definition of dissipation and its role in the RANS equations \label{sec-equations}}

\subsection{The instantaneous equations}
If we denote $\tilde{u_i}$ the instantaneous velocity components of a Newtonian incompressible fluid, $\rho$ the fluid  density, and $\mu$ the dynamic viscosity, then the flow is described for incompressible flow by the continuity equation and the Navier-Stokes equations given respectively by: 

\begin{eqnarray}
	\label{NS}
	\frac{\partial \tilde{u_i}}{\partial x_i}& =& 0\nonumber\\
	\rho\frac{\partial \tilde{u_i}}{\partial t} + \rho \tilde{u_j}\frac{\partial \tilde{u_i}}{\partial x_j}& =& -\frac{\partial \tilde{p}}{\partial x_i} +\frac{\partial }{\partial x_j}\left(\tilde{\tau}_{ij}\right)
\end{eqnarray}
For an incompressible Newtonian fluid, $\tilde{\tau}_{ij} = 2\mu \tilde{s}_{ij}$ is the viscous stress tensor, $\nu = \mu/\rho$ is the kinematic viscosity  and $\tilde{s}_{ij} = (1/2) (\partial \tilde{u}_i /\partial x_j  +  \partial\tilde{u}_j/\partial x_i)$ is the instantaneous strain-rate tensor, In these three papers only flows of constant density will be of interest.  We note for future reference that the deformation-rate tensor, $\partial \tilde{u}_i/\partial x_j$, can be decomposed into its symmetric and anti-symmetric parts as ${\partial \tilde{u}_i}/{\partial x_j} = \tilde{s}_{ij}+\tilde{\omega}_{ij}$
where $\tilde{\omega}_{ij} = (1/2) (\partial \tilde{u}_i /\partial x_j  -  \partial\tilde{u}_j/\partial x_i)$ is the rotation-rate tensor.

In turbulence the flow velocity field is usually decomposed into mean and fluctuating parts using the Reynolds decomposition, say $\tilde{u}_i = U_i + u_i$ , where $U_i = \langle \tilde{u}_i \rangle$ and the average of the fluctuating velocity, $\langle u_i \rangle = 0$. Any instantaneous flow variable (e.g., $\tilde{p}, \tilde{\tau} _{ij}, \tilde{\varepsilon} $...) can be similarly decomposed. Based on such a decomposition, it is possible to write a set of equations for the mean flow, the Reynolds-averaged equations given by:

\begin{eqnarray}
	\label{Reynolds}
	\frac{\partial U_i}{\partial x_i} &=& 0 \nonumber\\
	\rho\frac{\partial U_i}{\partial t} + \rho U_j\frac{\partial U_i}{\partial x_j} &=& -\frac{\partial P}{\partial x_i} +\frac{\partial }{\partial x_j}\left(\mathcal{T} _{ij} - \rho \langle u_iu_j\rangle\right)
\end{eqnarray}
where $\mathcal{T} _{ij} = 2\mu S_{ij} $, $S_{ij} = (1/2) (\partial U_i /\partial x_j  +  \partial U_j/\partial x_i)$ and  $\rho \langle u_iu_j\rangle$ are the well-known Reynolds stresses.

\subsection{The averaged equations}

The classical next step in a statistical approach of turbulence is to derive transport equations for this Reynolds stresses. To do so, first derive an equation for  the fluctuating velocity  by subtracting (\ref{Reynolds}) from (\ref{NS}). This equation is discussed in detail by \cite{george2014}. From this equation, it is possible,after to obtain the transport equations for the Reynolds stresses as:

\begin{eqnarray}
	\begin{gathered}
		\rho \frac{\partial \langle u_iu_j \rangle }{\partial t} + \rho U_l\frac{\partial \langle u_iu_j \rangle }{\partial x_l} = \\
		- \rho  \langle u_iu_l \rangle \frac{\partial U_j}{\partial x_l} - \rho  \langle u_ju_l \rangle\frac{\partial U_i}{\partial x_l}\\
		+ \frac{\partial}{\partial x_l}\left(\langle u_i\tau_{jl} \rangle + \langle u_j\tau_{il} \rangle - \langle u_i p \rangle\delta _{jl} - \langle u_j p \rangle\delta _{il} - \rho\langle u_i u_ju_l \rangle\right)\\
		+ \left(\langle p\frac{\partial u_i}{\partial x_j} \rangle + \langle p\frac{\partial u_j}{\partial x_i} \rangle\right)\\
		- \langle \tau_{il}\frac{\partial u_j}{\partial x_l} \rangle
		- \langle \tau_{jl}\frac{\partial u_i}{\partial x_l} \rangle
	\end{gathered}
	\label{rs_eq}
\end{eqnarray}
Although written slightly differently here from most papers and text books, this is still a well-known equation. The important point for this paper is to emphasize the form of the viscous diffusion (first two terms of the third line of equation (\ref{rs_eq}) and the dissipation (last line of the same equation). Note that the notion of ``dissipation'' appears here for the first time as the equations have raised by one order and the transported terms have now the dimension of a kinetic energy. As (\ref{rs_eq}) represents a set of 6 independent equations, the ``dissipation'' appears here as a tensor:

\begin{equation}
	\varepsilon _{ij}  =   \frac{1}{\rho}\langle\tau_{il}\frac{\partial u_j}{\partial x_l} \rangle
	+  \frac{1}{\rho}\langle\tau_{jl}\frac{\partial u_i}{\partial x_l} \rangle
\end{equation}

From equation (\ref{rs_eq}), by contraction of the two indices $i$ and $j$ (or by taking the trace of the set of equations), and dividing by 2, the equation for the turbulence kinetic energy $k$ is directly obtained as: 

\begin{equation}
	\rho\frac{\overline{D} k}{\overline{D} t} = \rho\langle u_iu_j \rangle\frac{\partial U_j}{\partial x_j} +   \frac{\partial}{\partial x_j} \left[-\langle pu_j \rangle_- \frac{\rho}{2}\langle u_iu_iu_j \rangle  +   \langle u_i \tau _{ij} \rangle \right] ~- ~ \rho  \varepsilon 
	\label{eq:k}
\end{equation}
where $\varepsilon \equiv \langle \tau_{ij} \partial u_i/\partial x_j \rangle = \frac{1}{2}\varepsilon_{ii}$ is the true turbulence dissipation,  $k \equiv \langle u_iu_i \rangle/2$ is the average kinetic energy per unit mass associated with turbulence, and the left-hand-side has been compacted using the averaged material derivative defined to be:

\begin{equation}
	\frac{\overline{D} }{\overline{D} t} = \frac{\partial }{\partial t} + \rho U_j\frac{\partial }{\partial x_j}. 
\end{equation}
Note that since ensemble averages are implied, the above equations (\ref{Reynolds}), (\ref{rs_eq}) and (\ref{eq:k}) are valid in both statistically stationary and non-stationary flows; i.e., there has been
no assumption of time-averaging!
Note also for future reference the form of the  viscous diffusion term, $\langle u_i\tau_{ij} \rangle$.


\subsection{The dissipation, $\varepsilon$}
For reasons of simplicity and to conform to usual practice, in the following we will replace $\langle \epsilon \rangle$ by just $\epsilon$.
It is relatively easy to show that $\langle \varepsilon \rangle$ is indeed the real dissipation (at least in a Newtonian flow), since it is always positive {\it and occurs with the opposite sign in the Reynolds-averaged entropy equations}.

It is easy to show from the definitions that for incompressible Newtonian fluids:
\begin{equation}
	\varepsilon = \langle \frac{1}{\rho}\tau_{ij}\frac{\partial u_i}{\partial x_j} \rangle  =2\nu\langle s_{ij}s_{ij} \rangle,
	\label{epsilon}
\end{equation}
since only the symmetrical part of the velocity deformation survives the double-contraction of the indices. Expanding the strain-rate tensor in equation~\ref{epsilon} yields the form most useful for experimental evaluation:
\begin{equation}
	\varepsilon = \nu\langle \frac{\partial u_i}{\partial x_j}^2 \rangle + \nu\langle \frac{\partial u_i}{\partial x_j}\frac{\partial u_j}{\partial x_i} \rangle
	\label{depsilon2}
\end{equation}

A source of considerable confusion in the turbulence literature is the second term on the right-hand-side of equation~\ref{depsilon2}.  If the flow is {\it statistically homogeneous}, then homogeniety alone implies that the upper and/or lower indices can be interchanged~\cite{george91}; e.g., permutting the lower yields:

\begin{equation}
	\langle \frac{\partial u_i}{\partial x_j} \frac{\partial u_j}{\partial x_i} \rangle = \langle \frac{\partial u_i}{\partial x_i} \frac{\partial u_j}{\partial x_j} \rangle
\end{equation}
If the flow is also incompressible, continuity together with homogeneity implies that cross-derivative terms are identically zero ALWAYS since $\partial u_i/\partial x_i = 0$ instantaneously.  But there are very few turbulent slows which are homogeneous.  So strictly speaking, this is almost never true, and is at best an approximation.  When it is true the flow is said to be ``locally homogeneous''~\cite{george91}.  part II \cite{george20} of this contribution will discuss in detail the degree to which our results can be considered to be locally homogeneous.


\subsection{The `pseudo-dissipation'}
There is another quantity which is often confused with the true dissipation, and  will also be investigated in the present contribution; namely:

\begin{equation}
	\mathcal{D} = \nu \langle \left[\frac{\partial u_i}{\partial x_j}\right] ^2 \rangle.
	\label{depsilon}
\end{equation}
Why this is problematical can easily be seen by expanding the double contraction of the velocity gradient tensor to yield:
\begin{equation}
	\mathcal{D} = \nu \langle \left[\frac{\partial u_i}{\partial x_j}\right] ^2 \rangle = \nu \langle [ s_{ij} + \omega_{ij}]^2 \rangle = \nu [\langle s_{ij} s_{ij} \rangle + \langle \omega_{ij} \omega_{ij} \rangle]
	\label{depsilon3}
\end{equation}
where $\omega_{ij}$ is the fluctuating rotation-rate tensor and the product
$s_{ij} \omega_{ij} = 0$ since $s_{ij}$ is symmetric and $\omega_{ij}$  antisymmetric.  Note that the mean square fluctuating rotation-rate tensor, $\langle\omega_{ij}\omega_{ij} \rangle$, is NOT in general equal to the mean square fluctuating strain-rate, $\langle s_{ij} s_{ij}\rangle$.  So in general $\varepsilon$ and $\mathcal{D}$ are different, even when the flow is turbulent.

It cannot be emphasized strongly enough that a distinction should be made between the true dissipation, $\varepsilon$, and the pseudo-dissipation, $\mathcal{D}$, which is defined from the mean square deformation rate.  Even though $\mathcal{D}$ has only squared terms, it is NOT the dissipation, since it is $\varepsilon$ which appears in the kinetic energy equation (\ref{eq:k}) and with opposite sign in the entropy transport equation (not detailed here).

\subsection{Why has $\varepsilon$ versus $\mathcal{D}$ been so confusing?}
That there is considerable confusion about $\varepsilon$ versus $\mathcal{D}$ in the turbulence literature is most likely for two reasons. First, as noted above, there is a particular circumstance in which $\varepsilon$ and $\mathcal{D}$ are equal, namely {\it homogeneous turbulence}.  The second reason is that in a homogeneous turbulence $\langle \omega_{ij}\omega_{ij} \rangle = \langle s_{ij}s_{ij} \rangle$ \cite{george91}.  But only in homogeneous turbulence! 

It is often assumed (usually without justification) that turbulence is {\it locally homogeneous}, which was  {\it defined} by \cite{george91} to be the situation where the enstrophy and mean square strain-rate are {\it approximately} equal, even if the flow is not globally homogeneous. This is usually justified by the same arguments used for the assumption of {\it local isotropy}; namely that the small scales which dominate velocity derivatives exist in an environment that is relatively independent of the large scale motions.  Sometimes this is approximately true, especially at high turbulence Reynolds numbers {\it and away from surfaces or stagnation points of the mean flow}.

The second reason for the confusion is that turbulence modelers prefer to use a form of the energy equations in which $\mathcal{D}$ (and not $\varepsilon$) occurs explicitly: 

\begin{equation}
	\frac{\overline{D} k}{\overline{D} t} = \langle u_iu_l \rangle\frac{\partial U_j}{\partial x_l} + \frac{1}{\rho} \frac{\partial}{\partial x_j} \left[-\langle pu_j \rangle - \frac{1}{2}\langle u_iu_iu_j \rangle   +   \mu \frac{\partial k}{\partial x_j} \right] ~-~\mathcal{D}
	\label{eq:Dversionenergy}
\end{equation}
Note that BOTH $\varepsilon$ and the viscous diffusion term $ \langle u_i\tau_{ij} \rangle$ on the right-hand-side of equation \ref{depsilon2} have been replaced by $\mathcal{D}$ and $\partial k/ \partial x_j$. 
This can be best understood by comparing the two forms of the energy equations, equations~\ref{eq:Dversionenergy} and \ref{eq:k}.  Since all the other terms in both equations are the same it follows immediately that:

\begin{equation}
	\frac{\partial}{\partial x_j}  \left[
	\nu \frac{\partial}{\partial x_j} \langle u_i u_i \rangle/2 \right] ~-~ \mathcal{D} ~ = ~ \frac{\partial}{\partial x_j} \left[ 2 \nu\langle u_i s_{ij} \rangle\right] ~ - ~\varepsilon
\end{equation}
But using the definitions of $\varepsilon$ and $s_{ij}$, the right-hand-side can be written as:

\begin{equation}
	\left\{\frac{\partial}{\partial x_j} \left[ 2 \nu\langle u_i s_{ij} \rangle\right] \right\} ~ - ~[\varepsilon]  = \left\{\frac{\partial}{\partial x_j} \langle u_i u_i \rangle/2  ~ + \nu \langle \frac{\partial u_i}{\partial x_j}\frac{\partial u_j}{\partial x_i}  \rangle \right\}~-~ \left[ \mathcal{D} ~ + ~ \nu \langle \frac{\partial u_i}{\partial x_j}\frac{\partial u_j}{\partial x_i}  \rangle \right]
\end{equation}
where the terms in curly and square brackets on the left-hand-side correspond to the expanded versions on the right-hand-side.
Thus equation (\ref{eq:Dversionenergy}) is exact and strictly equivalent mathematically to equation (\ref{eq:k}).  But with ``dissipation'' and ``diffusion'' terms that are not physically easy to interpret.  part II \cite{george20} of this contribution shows that this is a serious problem especially for $\varepsilon_{ik}$ inside of $y^+ = 30$, a fact seemingly only previously noticed by
\cite{jakirlic02}.

\section{A brief history of recent developments \label{app:asymptotic_review}}

The paper of Millikan~\cite{millikan38} and Clauser~\cite{Clauser56} represented the beginning of modern asymptotic analysis, and many have followed them, since (e.\ g.\, ~\cite{panton05,gadelhak94,afzal82,meneveau13}) to cite but a few, most following classical asymptotic analysis ideas or some version of Millikan's original analysis (c.f. \cite{TL72}).  The exception was
the approach of \cite{george97b,wosnik00} who introduced two new analytical techniques:  equilibrium similarity and Near-Asymptotics.  (\cite{george2006,george2007} provide a review of both ideas and their application.)  Unlike the earlier analyses which treated all wall-bounded flows the same, the new theories are different,
in large part because of the fact that some wall-bounded flows grow downstream (like boundary layer and wall jets), while others were statistically homogenous in the streamwise direction. Among the differences, the former gave power laws for the mean velocity profile, while the latter gave the familiar log laws. In inner variable these were written as $U/u_\tau = C_i {y^+}^\gamma$ and $U/u_\tau = {1/\kappa} \ln y^+ + B_i$ respectively. In both cases, the parameters were weak functions of the log of the local Reynolds number, $\ln \delta^+$, and only asymptotically constant, hence the term `Near-asymptotics'.  

Mathematically and physically the differences were entirely due to how the integral momentum was conserved in the streamwise direction: constant and determined by the balance between the pressure-gradient and wall-shear stress for channels and pipes; but continuing to evolve in the streamwise direction as reflected in the non-zero streamwise derivative of the  momentum thickness for evolving boundary layers.  The result for the zero-pressure gradient was that the outer region was scaled not by just one velocity parameter, $u_\tau$, but two, $u_\tau$ and $U_\infty$.  While the earlier Clauser analysis had to neglect terms of order $u_\tau/U_\infty$, a major inconsistency since $d\theta/dx = (u_\tau/U_\infty)^2$, the new theory for evolving flows was valid to second-order in $(u_\tau/U_\infty)$, consistent with a non-zero $d \theta/dx$.

Many new experiments at higher Reynolds numbers than before followed for both pipes and boundary layers, most claiming to confirm the log law but truth-to-tell, few (if any) made an honest attempt to evaluate the new results. Only \cite{george2006,george2007} tried to evaluate both theories in an objective manner, and interestingly both gave virtually identical results for the mean velocity profiles.  Even so the difference between log and powers over the range of Reynolds numbers explored was probably negligible anyway, so mostly were of theoretical interest only.  \cite{george97b,wosnik00} also provided for the first time deductions about the dissipation, and the present Part I of our contribution has shown them to be at least in general agreement with the data.

One mystery about the behavior of the turbulence intensities was resolved by the contribution of Hultmark~\cite{hultmark2012} using the same \cite{george97b} methodology. This was particularly important since it found  the missing log profile in the overlap region for the turbulence intensities; i.e., $k = A \ln y^+ +  F$, where $A$ and $F$ are weak functions of $\ln \delta^+$ and asymptotically constant. This had been long expected, since Townsend~\cite{townsend76} and Perry and Chung~\cite{perry82b}).  but never observed. Hultmark further related it to the apparent absence of the  ``inner peak'' in the previous turbulence intensity measurements, but which was observed inside the overlap region with the new turbulence intensity measurements at higher Reynolds number in Superpipe. The fact that this peak was missing at lower Reynolds number was the reason the log profile had not been observed.  The slope of the intensity was negative and only at higher Reynolds number had it emerged enough to make the log decay of the intensity profile visible.  The Reynolds number dependence of the inner peak was also strong support for the mesolayer argument of ~\cite{george97b,wosnik00}; namely that there existed a region in which the single point RANS equations were effectively inviscid, but the multi-point equations were not. Similar relations can be easily derived for the kinetic energy profiles for a developing boundary layer, but yielding power law relations instead; i.e.., $k =A_i {y^+}^\alpha$ instead.

As a final note, crucial to the arguments of \cite{george97b} and which determined that asymptotic value of the coefficients had to be constant was the assumption that K41 thinking applied to the overlap region in the limit of infinite Reynolds number. In particular, that the dissipation had to be finite in the limit of infinite Reynolds number.   If this constraint is relaxed, the asymptotic power goes to zero and the power law because a log law in the limit.  Recent investigations at Princeton \cite{vallikili15} and Lille \cite{george09} strongly suggest that K41 does not apply to the overlap region, but only to the outer flow (if at all).  To this point the analysis has not been modified to take that into account.

\end{document}